\documentclass[
a4paper,
aps,
prd,
preprint,
tightenlines,
preprintnumbers,
nofootinbib
]{revtex4-1}

\usepackage{fullpage}
\usepackage{amsfonts}
\usepackage{amsmath}
\usepackage{slashed}
\usepackage{amssymb}
\usepackage{graphicx}
\usepackage{epic}
\usepackage{eepic}
\usepackage{epsfig}
\usepackage{latexsym}
\usepackage{color}
\usepackage{float}
\usepackage{multirow}
\usepackage{hyperref}
\usepackage[caption=false]{subfig}
\usepackage{natbib}

\usepackage{shorthand}

\usepackage{mciteplus}

\newcommand{\ubr}[1]{\raisebox{1.5ex}{\hspace{#1ex}$\frown$\relax}}

\begin{document}

\title{Exclusion limits on a scalar decaying to photons and distinguishing its production mechanisms}

\author{Tanumoy Mandal}
\email{tanumoy.mandal@physics.uu.se}
\affiliation{Department of Physics and Astronomy, Uppsala University, Box 516, SE-751 20 Uppsala, Sweden}


\begin{abstract}
LHC run-II has a great potential to search for new resonances in the diphoton channel. Latest 13 TeV data 
already put stringent limits on the cross sections in the diphoton channel assuming the resonance is produced 
through the gluon-gluon fusion. Many beyond the Standard Model (SM) theories predict TeV-scale scalars, which 
copiously decay to diphotons. Apart from the gluon-gluon fusion production, these scalars can also be dominantly 
produced in other ways too at the LHC namely through the quark-quark fusion or the gauge boson fusions like 
the photon-photon, photon-$Z$, $WW$ or $ZZ$ fusions. In this paper we use an effective field theory approach 
where a heavy scalar can be produced in various ways and recast the latest ATLAS diphoton resonance search  
to put model-independent limits on its mass and effective couplings to the SM particles. If a new scalar is 
discovered at the LHC, it would be very important to identify its production mechanism in order to probe 
the nature of the underlying theory. We show that combining various kinematic variables in a multivariate 
analysis can be very powerful to distinguish different production mechanisms one from the other.
\end{abstract}



\maketitle

\section{Introduction}
\label{sec:intro}

There are numerous theoretical motivations to expect that the Standard Model (SM) 
is not the complete story and the scalar spectrum of a larger theory 
may be richer than to possess only one neutral scalar - the Higgs boson. 
From this expectation searches for new scalars are continuously being carried out at the 
Large Hadron Collider (LHC) in various channels. Along these directions, no confirmed hint has been found 
so far in any of these searches. Nevertheless, some anomalies in some of 
these searches have drawn significant attention in the high energy physics community, recently. 
Among them, the most famous one is the 750 GeV diphoton excess~\cite{Aaboud:2016tru,Khachatryan:2016hje} which created
a lot of excitements in the community. Before the excess went away with more data, 
numerous attempts have been made to explain the excess (see Ref.~\cite{Strumia:2016wys} for a
review and a long list of references). Another important excess was the diboson excess around 
2 TeV resonance mass~\cite{Aad:2015owa,Aad:2014xka,Aad:2015ufa} which also later turned out 
to be a statistical fluctuation.

Searches for heavy scalars at the LHC are generally being carried out in the diphoton, diboson or dijet 
resonance searches. The diphoton channel, among them, is particularly important as this channel
provides a comparatively cleaner background. Higgs boson was first discovered in the diphoton channel at the
LHC~\cite{Aad:2012tfa,Chatrchyan:2012xdj}. TeV-scale scalars decaying into a diphoton system is
one of the key predictions of many beyond the Standard Model (BSM) theories. Various possibilities
have been extensively explored in the context of the 750 GeV diphoton excess (see the reference list of~\cite{Strumia:2016wys}
for various models that predict TeV-scale diphoton resonances). To test these predictions, LHC run-II 
provides us a great opportunity to observe a diphoton resonance of mass up to a few TeV.
In this paper we particularly focus on the diphoton final state for these reasons.

If a particle decays to diphotons, it must either be spin-0 or spin-2 in nature as the
spin-1 particles decay to on-shell diphotons is forbidden by the Landau-Yang theorem \cite{Landau:1948kw,Yang:1950rg}.
A spin-2 particle or graviton couples universally to all matter fields through energy-momentum
tensor. Various extra dimensional models like the ADD model~\cite{ArkaniHamed:1998rs} or the RS model~\cite{Randall:1999ee}
predict the existence of graviton. If a resonance in the diphoton system mediated by graviton is observed, one
would expect resonances at same mass in other possible channels also. Therefore, simultaneous studies
in various channels might be more illuminating for the spin-2 particle. The current limits
on the graviton mass is already quite high, around $\sim 2-3$ TeV~\cite{Aaboud:2017yvp,CMS:2017xrr}. On the other hand,
scalars of mass $\sim 1$ TeV decaying into diphotons, which is a typical signature of many models,
are still allowed by the LHC data. These scalars can be produced at the LHC in various ways
{\it viz.} through the $gg$, $qq$, $\gm\gm$, $\gm Z$, $WW$ or $ZZ$ fusions. In this paper 
we consider a model-independent effective field theory (EFT) approach where the scalar can be produced 
and decayed (two-body) in different possible ways as mentioned. But we only concentrate on the diphoton 
decay mode in this paper as stated earlier.

First, we derive the available parameter space for a scalar (produced in different ways)
decays to diphotons using our EFT approach. These limits will be grossly model-independent and
can be used to set limits on other models wherever applicable. If a scalar resonance will actually
be seen in future, the most obvious question that will arise is - how the 
scalar is produced? A most common way to decipher the production mechanism of a heavy scalar is to look at
various kinematic distributions especially various jet observables, which are important in this regard. 
This has been investigated to some extent in the literature in the context of the 750 GeV diphoton 
excess~\cite{Csaki:2016raa,Harland-Lang:2016qjy,Ebert:2016idf,Dalchenko:2016dfa,Fuks:2016qkf}. In this
paper we revisit some of the jet observables and show their effectiveness in distinguishing 
different production modes. We, then, use a multivariate analysis (MVA) by combining many 
kinematic variables to distinguish different production modes more efficiently.

This paper is organized as follows: in Section~\ref{sec:lag}, we employ an effective 
Lagrangian for the scalar $\Phi$,
in Section~\ref{sec:pheno}, we discuss about the decays 
and various production modes of $\Phi$ at the LHC and derive exclusion 
limits on the mass and couplings from the latest diphoton resonance
search data. 
In the same section, we discuss how two different production modes of $\Phi$ can be distinguished
using a MVA analysis. Finally, we conclude in Section~\ref{sec:conclu}.

\section{Effective Lagrangian}
\label{sec:lag}

We consider an EFT approach where a heavy scalar $\Phi$ interacts with the SM gauge bosons through 
the dimension-5 operators and with the SM quarks through the dimension-4 operators. Assuming $\Phi$ is a 
CP-even real scalar, we employ the following effective Lagrangian,
\begin{align}
\label{eq:lag}
\mc{L} &\supset - \frac{\kp_{gg}}{4\Lm}~\Phi G^a_{\mu\nu}G^{a;\mu\nu}
-\frac{\kp_{\gm\gm}}{4\Lm}~\Phi A_{\mu\nu}A^{\mu\nu}
-\frac{\kp_{ZZ}}{4\Lm}~\Phi Z_{\mu\nu}Z^{\mu\nu}\nn\\
&-\frac{\kp_{\gm Z}}{2\Lm}~\Phi A_{\mu\nu}Z^{\mu\nu} 
-\frac{\kp_{WW}}{2\Lm}~\Phi (W^{+})_{\mu\nu}(W^{-})^{\mu\nu}
- \sum_q \frac{\kp_{qq}v}{\Lm}~\Phi \bar{q}q\ ,
\end{align}
where the field-strength tensors corresponding to gluon ($g$), photon ($\gm$), $W^{\pm}$ and $Z$ bosons 
are $G^a_{\mu\nu}$, $A_{\mu\nu}$, $(W^{\pm})_{\mu\nu}$ and $Z_{\mu\nu}$ respectively, 
and their generic form is, $\mc{V}_{\mu\nu}=\partial_{\mu}\mc{V}_{\nu}-\partial_{\nu}\mc{V}_{\mu}$
where $\mc{V}=\{g,\gm,W^{\pm},Z\}$. 
All the dimension-5 operators are suppressed by the new physics scale $\Lm$. In general,
$\Lm$ could be different for different operators, but we assume they are same
for all the operators. Note that only the $\Phi\bar{q}q$ is a dimension-4 operator and we introduce the electroweak
symmetry breaking scale
$v\approx 246$ GeV in association with $\kp_{qq}$ to bring the scale $\Lm$ in the interaction. The motivation behind
this is that if $\Phi\bar{q}q$ operators are effectively originated from new physics then effective couplings are expected to contain the imprint of the scale $\Lambda$ (possibly in the form $v/\Lambda$ or some power of this ratio).
This way of parameterizing the $\Phi\bar{q}q$ couplings
also enables us to present exclusion limits on all the couplings in the $\kappa_{xy}/\Lambda$ form.
Here, we use the notation $\kp_{xy}$ to denote a generic dimensionless coupling
associated with the $\Phi xy$ vertex.
The scalar $\Phi$ can, in general, couple differently with the different
SM quarks. For simplicity, in this analysis, we assume a single coupling $\kp_{qq}$ same for
all the SM quarks. Note that in all interactions with the gauge bosons, the normalization 
factor are so chosen such that the corresponding Feynman rule takes the form,
\begin{align}
(i\kp_{xy}/\Lm)\lt(g^{\al\bt}p_1\cdot p_2 - p_1^{\bt}p_2^{\al}\rt)\ ,
\end{align}
where $p_1$ and $p_2$ are the 4-momenta of two gauge bosons $\mc{V}_1^{\al}$ and 
$\mc{V}_2^{\bt}$ respectively directed towards the vertex. The Feynman rule for the
$\Phi\bar{q}q$ interaction is $i\kp_{qq}v/\Lm$.

In general, the new scalar $\Phi$ can mix with the 125 GeV scalar ($h_{125}$) with a mixing angle $\al$. This
leads to the scaling of all the couplings of $h_{125}$ by a factor $\cos\al$. Although this would not change the
branching ratios (BRs) of $h_{125}$, it would change the production cross section of $h_{125}$ by a factor $\cos^2\al$. Since all
the measured signal strengths are pretty close to unity, this will make $\cos\al$ close to one. 
That is why, in this paper we have neglected any mixing between $\Phi$ and $h_{125}$ for simplicity.

\section{Phenomenology}
\label{sec:pheno}

In addition to the SM Lagrangian, we implement the effective Lagrangian of $\Phi$ shown in
Eq.~\eqref{eq:lag} in {\sc FeynRules}~\cite{Alloul:2013bka} to generate the 	
Universal FeynRules Output~\cite{Degrande:2011ua} model files for the 
{\sc MadGraph}~\cite{Alwall:2014hca} event generator. We use the 
MMHT14LO~\cite{Harland-Lang:2014zoa} parton distribution functions (PDFs) 
for event generation.
This PDF set includes the photon PDF which has been computed following the approach described
in~\cite{Harland-Lang:2016qjy,Harland-Lang:2016apc}. We use the factorization scale $\mu_F$ and
the renormalization scale $\mu_R$ at $M_{\Phi}$ in our analysis. Generated events are further showered and
hadronized including multiple parton interactions by using {\sc Pythia8}~\cite{Sjostrand:2007gs}. 
We perform detector simulation using {\sc Delphes}~\cite{deFavereau:2013fsa} which uses {\sc FastJet}~\cite{Cacciari:2011ma} for jet clustering. Jets are clustered using the 
anti-$k_T$ algorithm~\cite{Cacciari:2008gp} with $R=0.4$. 
We analyze the reconstructed objects by implementing ATLAS selection cuts~\cite{Aaboud:2017yyg}, which we summarize 
in~\ref{ssec:exclu}.
For MVA, we use the adaptive Boosted Decision Tree (BDT) algorithm in 
the TMVA~\cite{Hocker:2007ht} framework.

\subsection{Decays of $\Phi$}

From the Lagrangian in Eq.~\eqref{eq:lag}, we have the following two-body decay modes of $\Phi$ 
\emph{viz.} $\Phi\to xy$ where $xy=\{gg,qq,\gm\gm,\gm Z,WW,ZZ\}$. The partial widths for these decay modes
are given by the following expressions,
\begin{align}
\label{eq:pwformula}
\Gm_{gg}&=\frac{\kp_{gg}^2M_{\Phi}^3}{8\pi\Lm^2};~~~
\Gm_{qq}=\frac{3\kp_{qq}^2v^2M_{\Phi}}{8\pi\Lm^2}\lt(1-\frac{M_q^2}{M_{\Phi}^2}\rt)^{\frac{3}{2}};~~~
\Gm_{\gm\gm}=\frac{\kp_{\gm\gm}^2M_{\Phi}^3}{64\pi\Lm^2};\nn\\
\Gm_{\gm Z}&=\frac{\kp_{\gm Z}^2M_{\Phi}^3}{32\pi\Lm^2}\lt(1-\frac{M_Z^2}{M_{\Phi}^2}\rt)^3;~~~
\Gm_{VV}=\frac{\kp_{VV}^2M_{\Phi}^3}{32\pi\Lm^2}\lt(1-\frac{M_V^2}{M_{\Phi}^2}\rt)^{\frac{1}{2}}
\lt(1 - \frac{4M_V^2}{M_{\Phi}^2} + \frac{6M_V^4}{M_{\Phi}^4}\rt)\ ,
\end{align}
where $V$ denotes the electroweak gauge bosons $W^{\pm}$ and $Z$. There could be subdominant three-body 
decays of $\Phi$ possible mediated through 
an off-shell gauge boson. If the intermediate gauge boson is massless, in case for gluons 
or photons, the three-body BRs are non-negligible especially
when $M_{\Phi}$ is large~\cite{Danielsson:2016nyy}. In this analysis, we consider the 
two-body and three-body decays of $\Phi$ to obtain the total width where the three-body decay widths
are computed numerically using {\sc MadGraph}. Partial widths of three-body decay modes where an
off-shell gauge boson goes to $W^+W^-$ pair grow very rapidly with increasing scalar mass.
This is due to the contribution coming from the longitudinal polarizations of $W$ bosons.
Therefore, in high mass region, BR for $\Phi\to\gm\gm$ reduces substantially.

\subsection{Production of $\Phi$ at the LHC}

\begin{figure}[h!]
\includegraphics[scale=0.6]{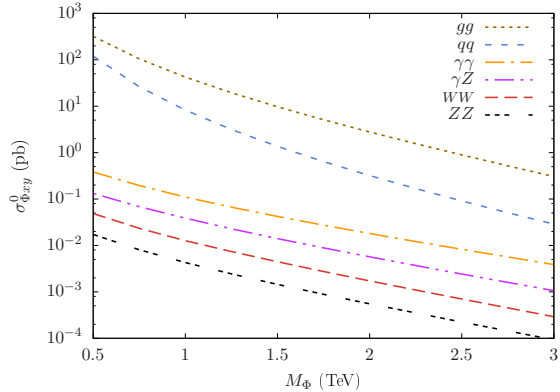}
\caption{Partonic cross sections of various production modes of $\Phi$ as functions of $M_{\Phi}$ 
computed at $\mu_R=\mu_F=M_{\Phi}$ 
at the 13 TeV LHC. Here, $\sg^0_{\Phi xy}$
denotes the cross section of $\Phi$ produced through the $xy$ fusion with $\kp_{xy}=1$ and $\Lm=1$ TeV. Initial $g$, $q$
and $\gm$ come from the PDFs of proton whereas initial $W$ and $Z$ come from initial quarks. These cross
sections are computed by applying some basic generation level cuts as defined in Eq.~\eqref{eq:cut}.}
\label{fig:cs}
\end{figure}
When all $\kp_{xy}$ in Eq.~\eqref{eq:lag} are nonzero, the scalar can be produced from 
the $gg$, $qq$, $\gm\gm$, $\gm Z$, $WW$ and $ZZ$ fusions at the LHC. 
In Fig.~\ref{fig:cs}, we show the partonic cross sections of different production modes of $\Phi$
at the 13 TeV LHC for $\kp_{xy}=1$ (taking one at a time) and $\Lm=1$ TeV. 
In case for the production of $\Phi$ through the $WW$ or $ZZ$ fusions, initial $W$ and $Z$ come from the quark 
splitting. Therefore, $\Phi$ is produced in association with at least two jets for this case.
Similarly, for the $\gm Z$ initiated production, $\Phi$ is produced in association with at least one jet. 
Partonic cross sections are computed by applying the following generation level cuts on the jets ($j$) and photons ($\gm$) 
wherever applicable
\be
\label{eq:cut}
p_T(j),p_T(\gm)>20~{\rm GeV};~|\eta(j)|<5,|\eta(\gm)|<2.5;~\Dl R(jj),\Dl R(\gm\gm),\Dl R(j\gm)>0.4\ .
\ee
Here, transverse momentum, pseudorapidity and separation in the $\eta-\phi$ plane are denoted by $p_T$, $\eta$ and 
$\Dl R$ respectively. These basic cuts are used to avoid any soft divergence present at the event generation level and 
stricter selection cuts are applied at the level of reconstructed event analysis after detector simulation. 
Note that all cross sections scale as $(\kp_{xy}/\Lm)^2$, and therefore, we present them by choosing
$k_{xy}=1$ and $\Lm=1$ TeV such that one can translate it easily for other values.

For all the six types of production of $\Phi$, we generate parton level events with up to two jets in the final state. These events are passed to {\sc Pythia8}~\cite{Sjostrand:2007gs}for showering and hadronization. This process may introduce double counting between the matrix element
partons and the parton showers. To generate inclusive signal events by avoiding any double counting, we use the
MLM matching~\cite{Mangano:2006rw} technique to match the matrix element partons with the parton shower. 
Inclusive signal events including up to two jets for the $gg$, $qq$ and $\gm\gm$ fusions are generated by combining the following processes,
\be
\left. \begin{array}{lclcl}
pp &\to &(\Phi) 		&\to & \gm\gm\ubr{-2.5}\,,\\
pp &\to &(\Phi\ j)		&\to & \gm\gm\ubr{-2.5}\ j\,,\label{eq:match}\\
pp &\to &(\Phi\ jj)  	&\to & \gm\gm\ubr{-2.5}\ jj\, ,
\end{array}\right\}
\ee
where we set the matching scale $Q_{cut} \sim 125$ GeV. The curved connections above two photons
signify that they come from the decay of $\Phi$. To determine the appropriate $Q_{cut}$
for these production processes, we have done three important checks {\it viz.} smooth transition in the 
differential jet-rate distributions between events with 
$N$ and $N+1$ jets, matched cross sections are within $\sim 10\%$ of the zero jet contribution and
also do not vary much with the $Q_{cut}$ variation once we have chosen it properly.
For the $\gm Z$, $WW$ or $ZZ$ fusion productions, the initial $W$ or $Z$ come from the quark splitting and we
have additional jets at the Born level process. Therefore, the $WW$ and $ZZ$ fusion events are generated
only at the $pp\to\Phi jj$ level and no matching is required for these cases. But for the $\gm Z$ fusion,
we do use matching by combining the processes $\gm p\to\Phi j$ and $pp\to\Phi jj$ with $Q_{cut} \sim 30$ GeV.
The dominant SM $\gm\gm$ background (about 90\% of the total) comes from the $q\bar{q}\to\gm\gm$
process. We generate this background by matching up to 2 jets with $Q_{cut} \sim 20$ GeV.


\subsection{Exclusion from the LHC data}
\label{ssec:exclu}

Diphoton resonance searches at the LHC using run-I and run-II data set strong upper limits (ULs) 
on $\sg\times{\rm BR}$ of a spin-0 or spin-2 resonances~\cite{Khachatryan:2016yec,Aaboud:2017yyg}. 
It should be noted that these searches are generally optimized for an $s$-channel resonance production 
through the $gg$ fusion followed by its decay to two photons. If the resonance is not produced from the
$gg$ fusion, the selection cut efficiencies can vary depending on the different production 
mechanisms of the resonance. For a particular production mechanism, it can also vary significantly on the number of 
selected photons and jets. Therefore, in order 
to derive exclusion limits on the model parameters by recasting the limits on $\sg\times{\rm BR}$ from 
an experiment, one has to properly take care of the selection cut efficiencies. This can 
be done properly by using the following relation~\cite{Danielsson:2016nyy}:
\be
\label{eq:Ns}
\mc{N}_s =(\sg\times {\rm BR})_s \times\ep_s\times \mc{L}=\sum_i (\sg\times {\rm BR})_i\times\ep_i\times\mc{L}\ ,
\ee
where $\mc{N}_s$ is the UL on the number of signal events, which can be written as the product
of the signal cross section $(\sg\times {\rm BR})_s$ (produced through a particular mechanism used in the analysis), the corresponding signal
cut efficiency $\ep_s$ and the luminosity $\mc{L}$. When different types of production mechanisms 
contribute to any experiment, $\mc{N}_s$ can be expressed by the sum
$\sum_i (\sg\times {\rm BR})_i\times\ep_i\times\mc{L}$. Here, $i$ runs over all the contributing
production mechanisms.

To see the change in efficiency for the different production
mechanisms and also for the different resonance masses, we 
roughly employ the following event selection cuts used by the ATLAS collaboration for their 
spin-0 diphoton resonance search as listed below~\cite{Aaboud:2017yyg}.
\ben
\item Transverse energy of the two selected photons satisfy $E_T(\gm_1)>40$ GeV and $E_T(\gm_2)>30$ GeV
and transverse momenta of selected jets satisfy $p_T(j)>25$ GeV for $|\eta(j)|<2.5$ and
$p_T(j)>50$ GeV for $|\eta(j)|>2.5$. Here, $\gm_1$ and $\gm_2$
denote the highest-$p_T$ and second highest-$p_T$ photons respectively.  

\item Pseudorapidity of the selected photons satisfy $|\eta(\gm)|<2.37$ excluding
the barrel-endcap region $1.37 < |\eta(\gm)| < 1.52$ and jets $|\eta(j)|<4.4$.

\begin{figure}[!htbp]
\includegraphics[scale=0.6]{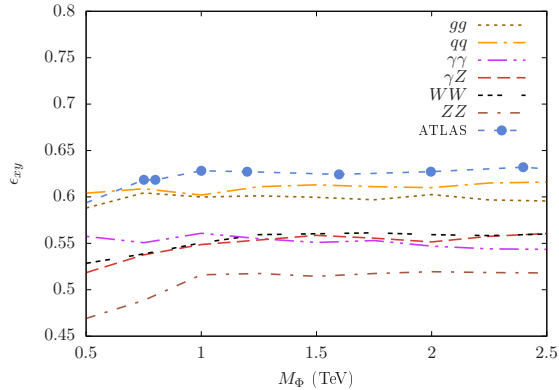}
\caption{Cut efficiencies for different production modes as functions of $M_{\phi}$ for the
ATLAS selection cuts as defined in the text.
The ATLAS cut efficiency (blue dots) is for a spin-0 resonance produced through the $gg$ fusion~\cite{Aaboud:2017yyg}.}
\label{fig:ce}
\end{figure}

\item Separation in the $\eta$-$\phi$ plane between the two selected photons or any photon-jet or jet-jet pair
satisfy $\Dl R(\gm_1,\gm_2),\Dl R(\gm_1,j),\Dl R(\gm_2,j),\Dl R(j,j) > 0.4$.
\item Invariant mass of the two selected photons $M(\gm_1,\gm_2)$ satisfies $E_T(\gm_1) > 0.4M(\gm_1,\gm_2)$
and $E_T(\gm_2) > 0.3M(\gm_1,\gm_2)$.
\een
In addition to the above set of cuts, we also apply default photon isolation cuts given in {\sc Delphes}
for the ATLAS detector. Jets with high-$\eta$ mainly come from the vector boson fusion topologies. We use
a threshold $p_T(j)>50$ GeV for $|\eta(j)|>2.5$ for better sensitivity~\cite{Kruse:2014pya}.
In Fig.~\ref{fig:ce}, we show cut efficiencies for the cuts listed above 
for different production modes as functions of $M_{\phi}$. The cut efficiency for ATLAS for their spin-0 resonance
produced through the $gg$ fusion is roughly about 62\%~\cite{Aaboud:2017yyg} and we find very close agreement (around 60\%) 
using our analysis codes. After validating our codes, we compute cut efficiencies for the other 
production modes for the selection cuts mentioned above and find that they do not vary much, only up to $\sim 15\%$ for different production modes. It is 
pointed out in the ATLAS paper~\cite{Aaboud:2017yyg} that the cut efficiencies for different production modes would not
differ much for their signal criteria (fiducial region).
As expected, in the high mass region, $M_{\Phi}\gtrsim 1$ TeV, cut efficiencies become insensitive to the mass.

\begin{figure}[!htbp]
\includegraphics[height=6cm,width=7cm]{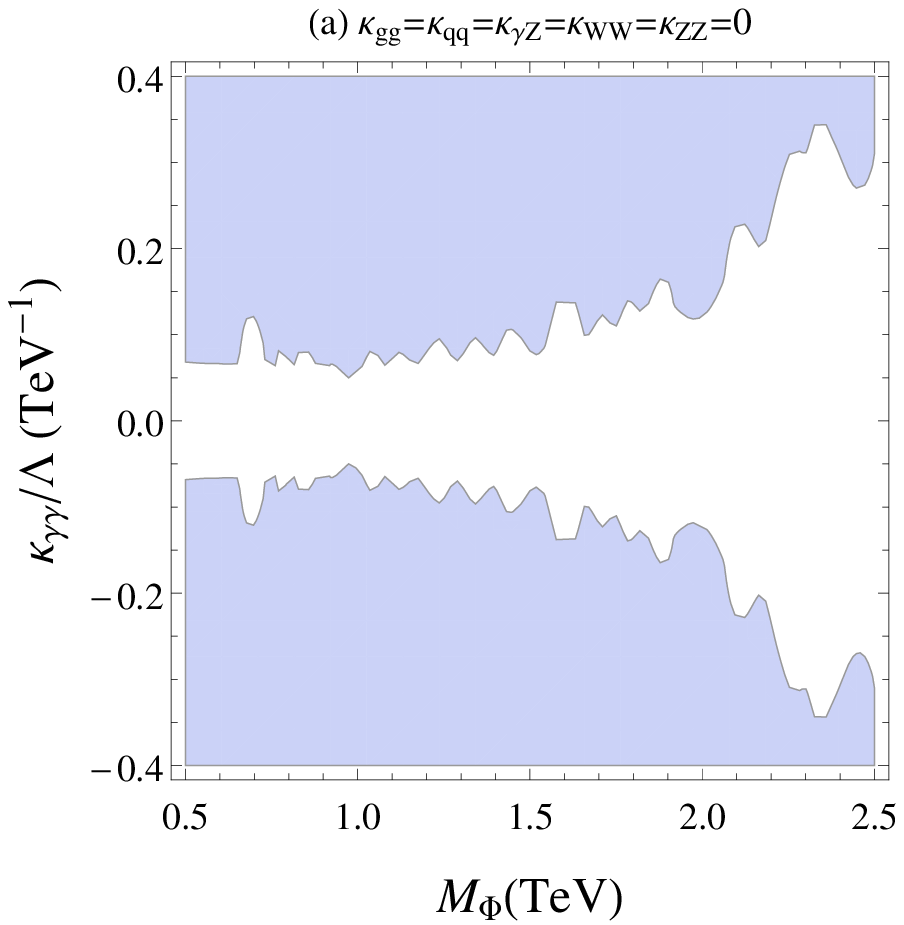}\hspace{0.25cm}
\includegraphics[height=6cm,width=7cm]{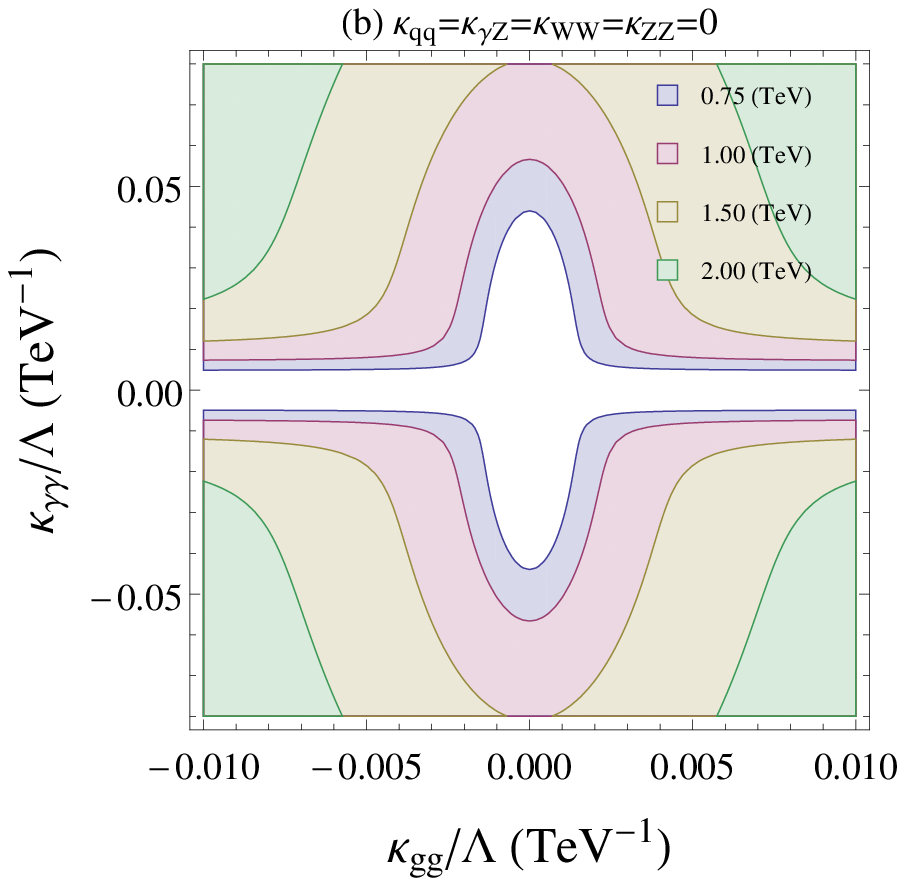}\\
\includegraphics[height=6cm,width=7cm]{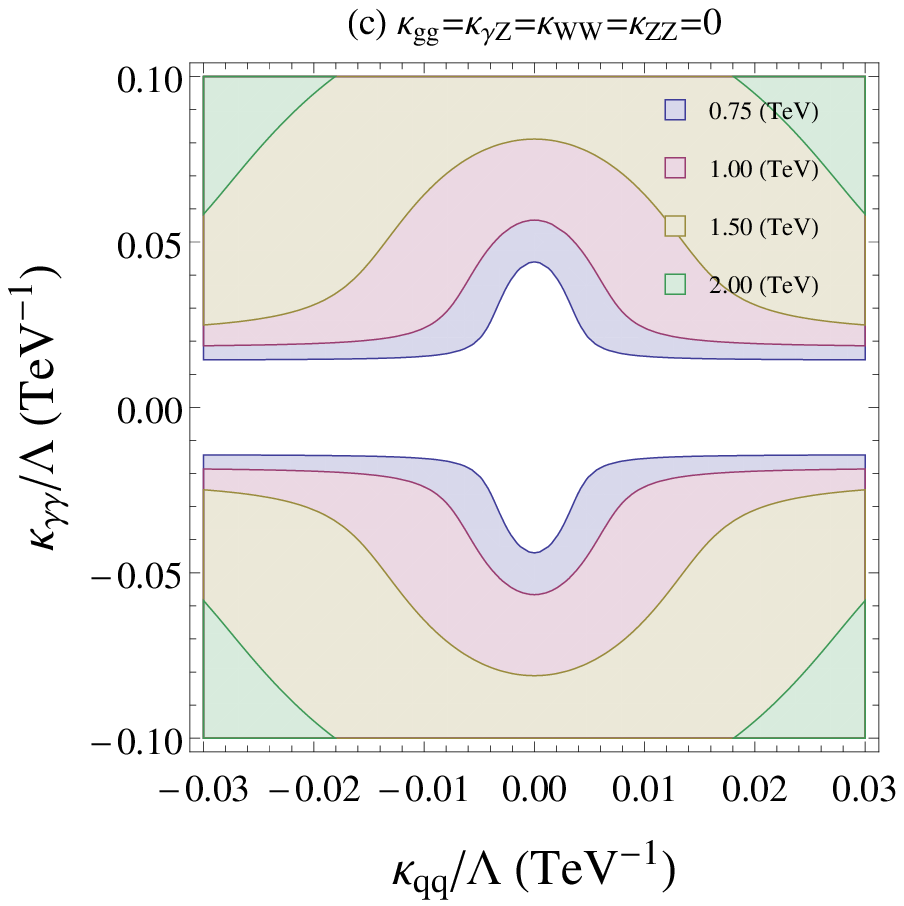}\hspace{0.25cm}
\includegraphics[height=6cm,width=7cm]{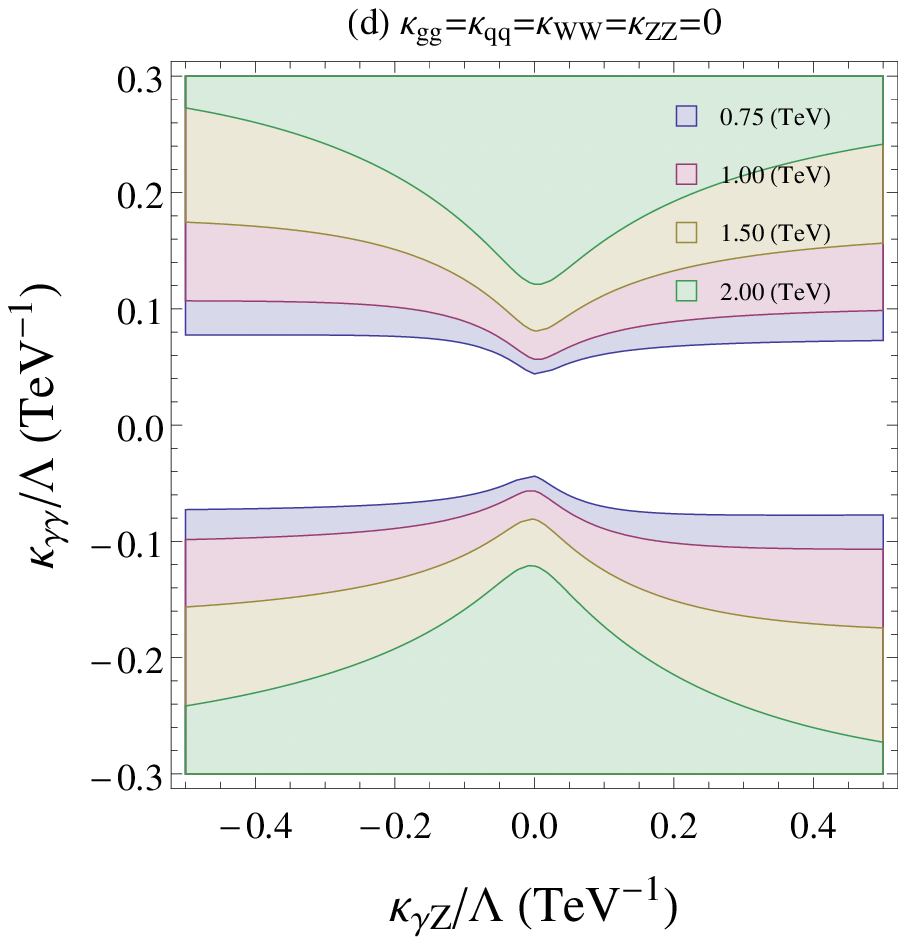}\\
\includegraphics[height=6cm,width=7cm]{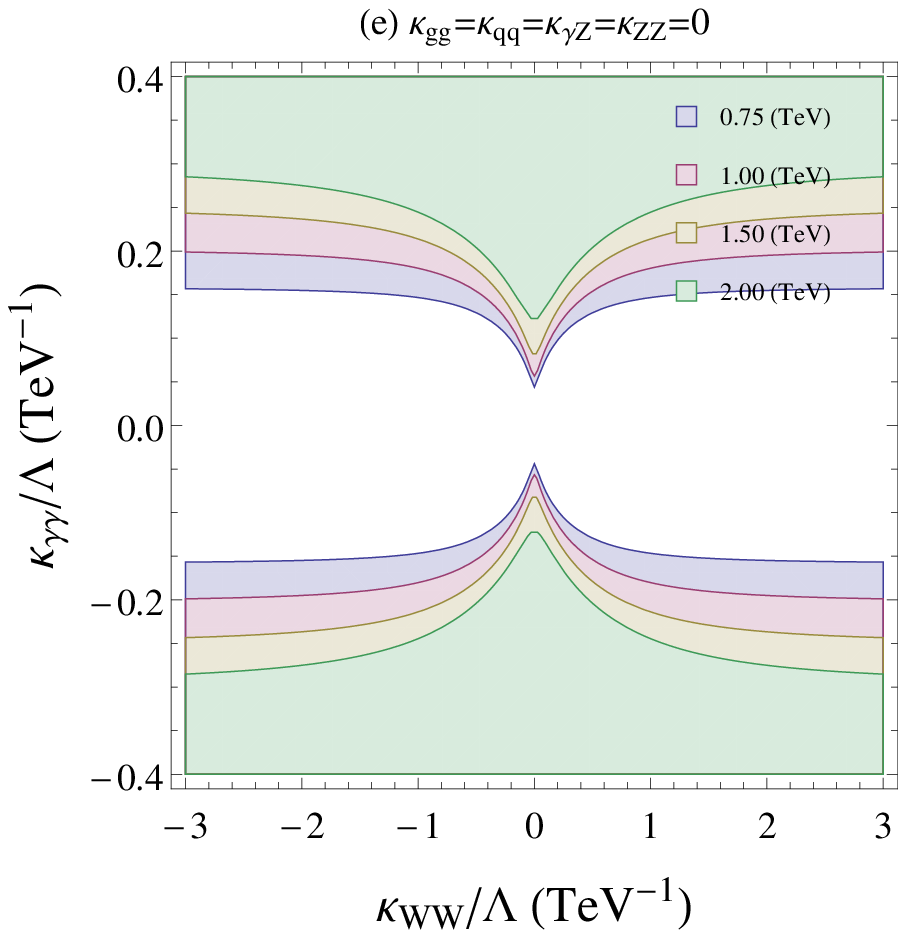}\hspace{0.25cm}
\includegraphics[height=6cm,width=7cm]{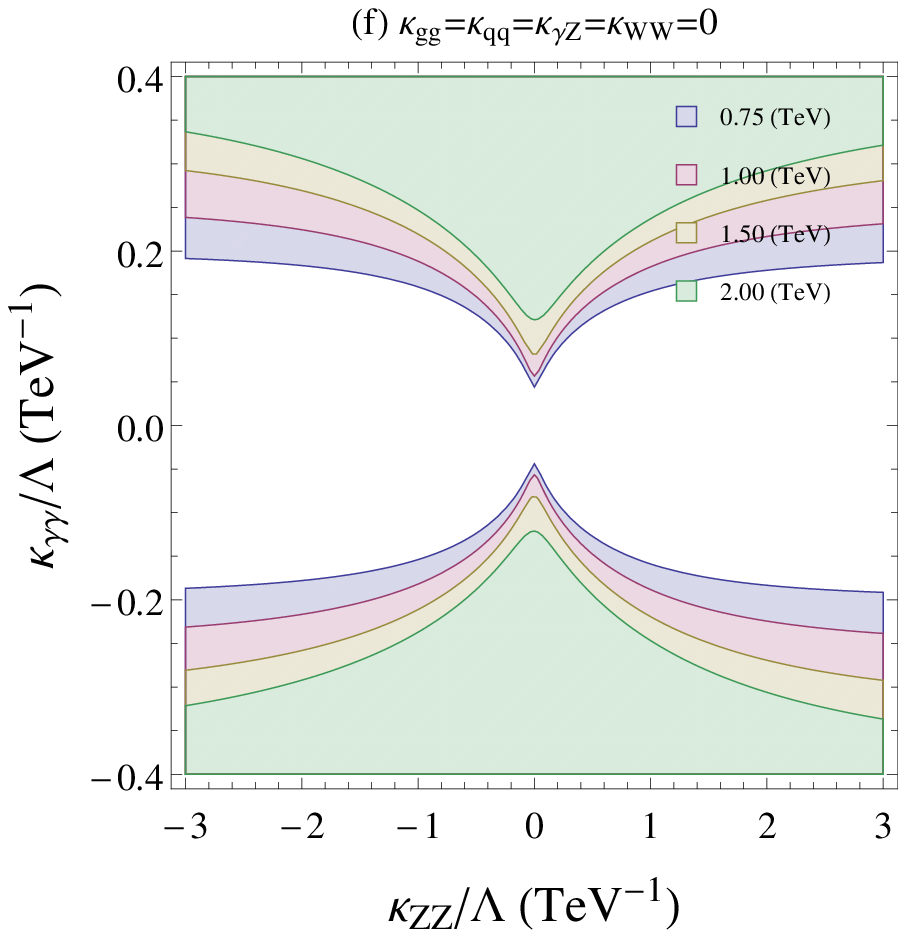}
\caption{Exclusion regions using the observed upper limits (with 95\% confidence level) on $\sg\times{\rm BR}$ 
for the spin-0 diphoton resonance search by ATLAS~\cite{Aaboud:2017yyg} ($\sqrt{s}=13$ TeV, $\mc{L}=36.7$ fb$^{-1}$). 
(a) Exclusions in the $M_{\Phi}-\kp_{\gm\gm}/\Lm$ plane setting all other $\kp_{xy}=0$. (b)-(f) Exclusions in the 
$\kp_{xy}/\Lm-\kp_{\gm\gm}/\Lm$ plane while setting all other $\kp_{xy}$ to zero for $M_{\Phi}=0.75,1,1.5,2$ TeV.}
\label{fig:exclusion}
\end{figure}

In our EFT approach, there are six free couplings $\kp_{xy}$ that affects the production of $\Phi$.
But taking all $\kp_{xy}$ nonzero at the same time will make the analysis very complicated.
Therefore, for simplicity, we choose only one $\kp_{xy}$ as nonzero at a time, in addition to 
nonzero $\kp_{\gm\gm}$, and show the two-dimensional (2D) exclusion regions (colored) in the $\kp_{xy}/\Lm-\kp_{\gm\gm}/\Lm$ plane for four 
benchmark masses, $M_{\Phi}=0.75,1,1.5,2$ TeV (presented in Fig.~\ref{fig:exclusion}). Only in Fig.~\ref{fig:exclusion}(a), we show the
exclusion regions (colored) in the
$M_{\Phi}-\kp_{\gm\gm}/\Lm$ plane assuming all $\kp_{xy}$ are zero except $\kp_{\gm\gm}$. 
To derive these limits, we recast 95\% confidence level (CL) UL on the $\sg\times\mathrm{BR}$ 
for the $\gm\gm$ spin-0 resonance search by the ATLAS collaboration at the 13 TeV with 
$\mc{L}=36.7$ fb$^{-1}$~\cite{Aaboud:2017yyg}. This analysis is done for the resonance width
$\Gm=4$ MeV. If the width of a particle is very small 
compared to its mass, one can safely
use the narrow width approximation (NWA). In all our results we use the NWA ignoring any interference effect
between the signal and the background.  

The bumpy nature in the exclusion limit on $\kappa_{\gamma\gamma}/\Lm$ in Fig.~\ref{fig:exclusion}(a) is due to  
non-smooth nature of the
observed UL on $\sg\times{\rm BR}$.
The highest value of 
$\kp_{\gm\gm}/\Lm$ that is excluded is $\sim 0.05$ around $M_{\Phi}\sim 1$ TeV. By choosing a value
for $\kp_{\gm\gm}$, one can translate this limit to $\Lm$. For instance, choosing $\kp_{\gm\gm}=1$ for 
$M_{\Phi}\sim 1$ TeV, one finds that $\Lm \lesssim 20$ TeV is excluded.
Basic shape of the exclusion regions in Figs.~\ref{fig:exclusion}(b) and~\ref{fig:exclusion}(c) are 
different from the ones in Figs.~\ref{fig:exclusion}(d), \ref{fig:exclusion}(e) and ~\ref{fig:exclusion}(f). This is because
the $gg$ and the $qq$ fusion productions dominate over the $\gm\gm$ fusion mode for $\kp_{xy}=1$ as seen
in Fig.~\ref{fig:cs}. On the other hand, cross sections for the $WW$, $ZZ$ and $\gm Z$ fusion modes
are smaller than the $\gm\gm$ mode for unity $\kp_{xy}$. One should also notice that exclusion 
regions in Figs.~\ref{fig:exclusion}(d) and~\ref{fig:exclusion}(f) are slightly asymmetric around $\kp_{\gm\gm}=0$
axis. This is due to the interference effect between the $\gm\gm$ and the $\gm Z$ or $ZZ$ production modes.
On the other hand, no interference is possible between the $\gm\gm$ and the $gg$, $qq$ or $WW$ fusion
modes.
Note that exclusion limits become insensitive to $\kappa_{xy}/\Lm$ as we go to higher values. This is because
the production cross section $\sg$ varies as $\kappa_{xy}^2$ and BR to diphoton
$\textrm{BR}_{\gm\gm}$ varies as $\sim \kappa_{\gamma\gamma}^2/\kappa_{xy}^2$ for $\kappa_{xy} \gg \kappa_{\gamma\gamma}$ region. 
This makes $\sigma\times\mathrm{BR}\sim \kappa_{\gamma\gamma}^2$ for large $\kp_{xy}$. 

To include higher order effects, we use a constant next-to-leading order (NLO) $K$-factor of 2 for the $gg$ fusion~\cite{deFlorian:2016spz}.
The NLO corrections to a heavy scalar produced from the $b\bar{b}$ fusion is computed in~\cite{Balazs:1998sb} 
where it is found that the NLO $K$-factor is close to 1 for heavier masses. If the scalar is produced
from the light quark fusions, one might expect slightly bigger $K$-factor. Here, we assume it to be 1
since it is not available in the literature. 
For the $\gm\gm$, $\gm Z$, $WW$ and $ZZ$ we assume it to be 1.3~\cite{Arnold:2008rz}. The actual
values of the $K$-factors for different channels can be slightly different from the constant values we 
have used but they have very little effect on the exclusion limits. 

\begin{figure}[!htbp]
\includegraphics[scale=0.65]{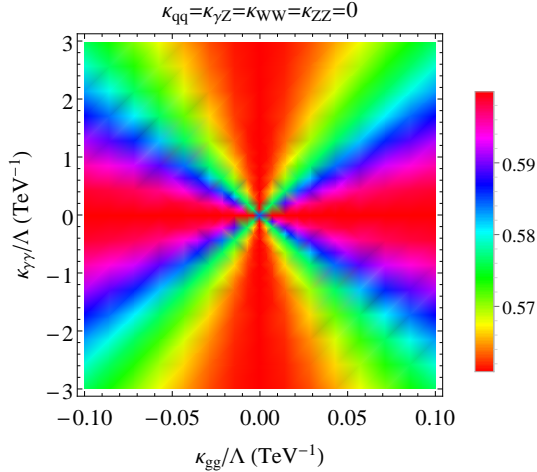}
\caption{Combined cut efficiency in the $\kp_{gg}/\Lm-\kp_{\gm\gm}/\Lm$ plane (all other $\kp_{xy}$ are zero) 
for $M_{\Phi}=1$ TeV for the ATLAS selection cuts as defined in the text.}
\label{fig:cecom}
\end{figure}

\begin{figure}[!htbp]
\includegraphics[width=0.3\textwidth]{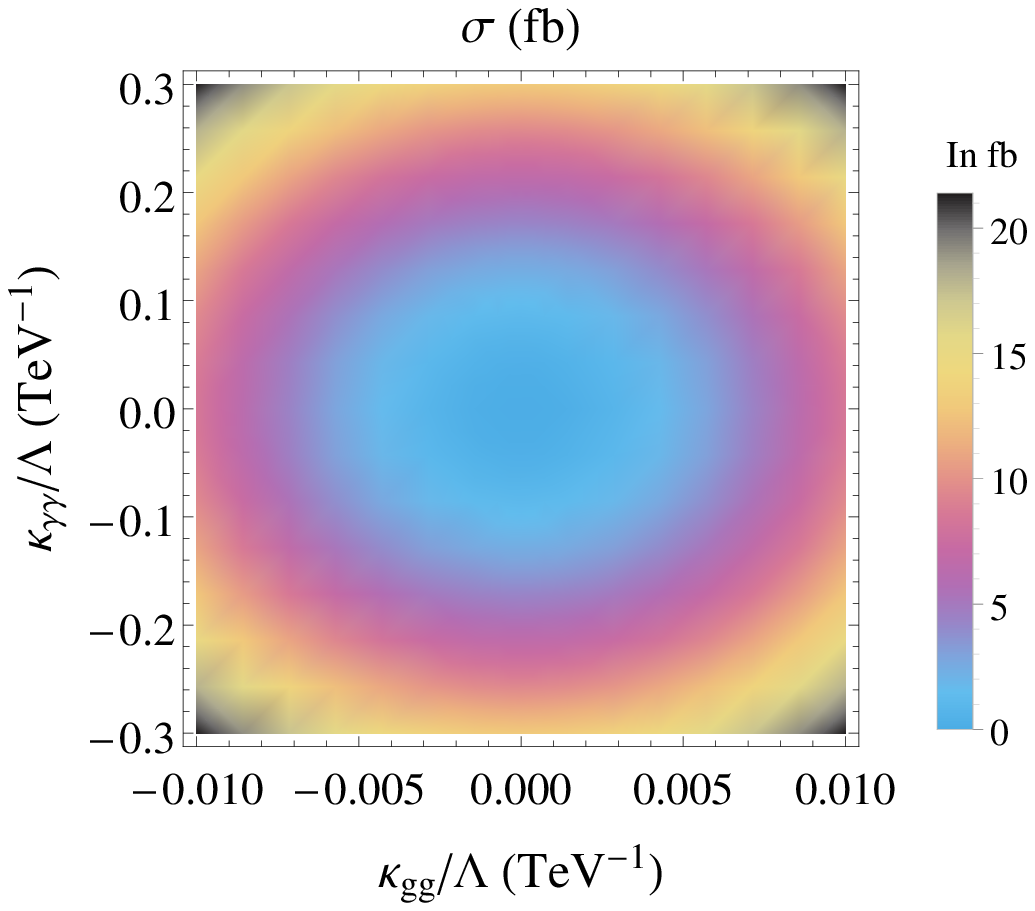}
\includegraphics[width=0.3\textwidth]{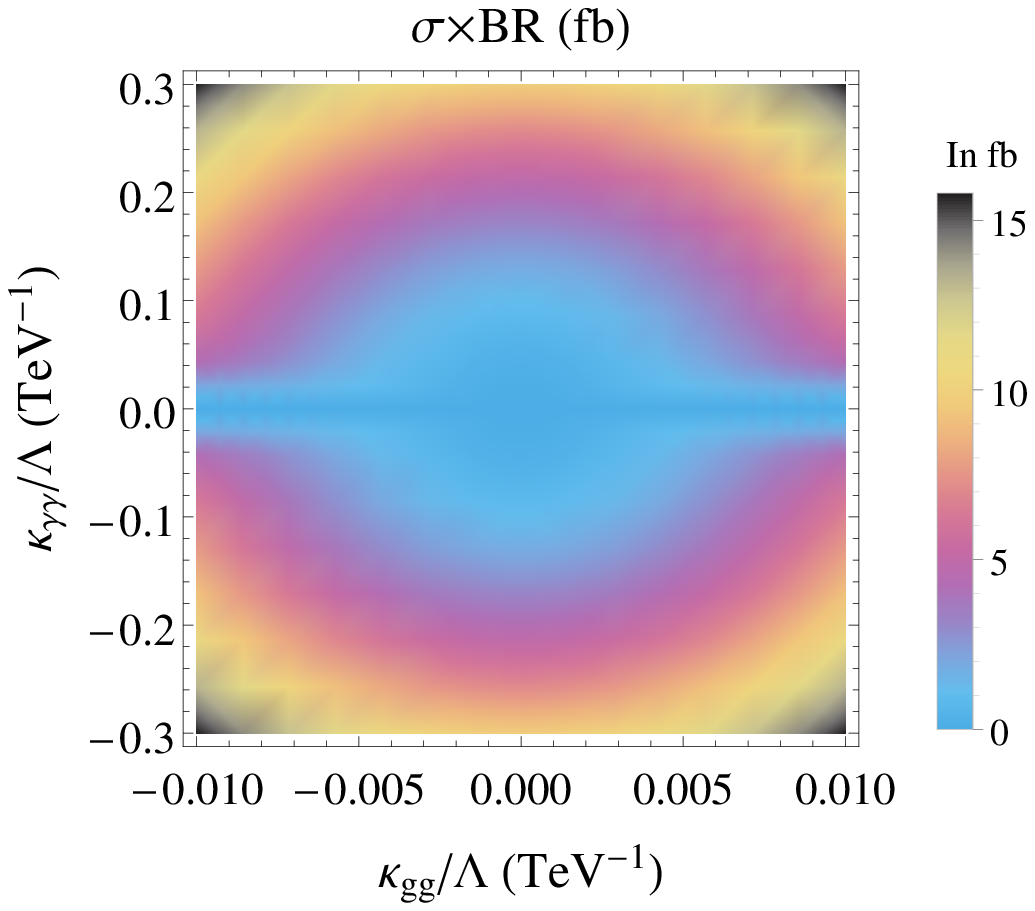}
\includegraphics[width=0.3\textwidth]{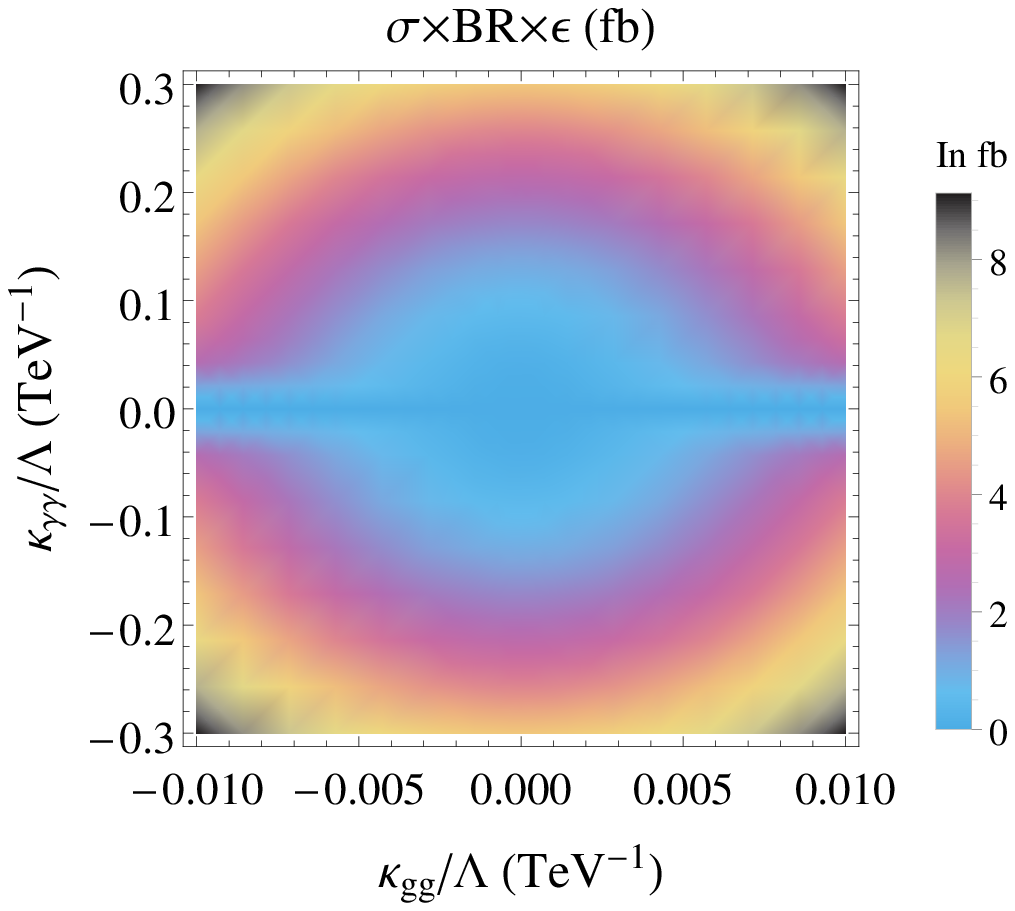}\\
\includegraphics[width=0.3\textwidth]{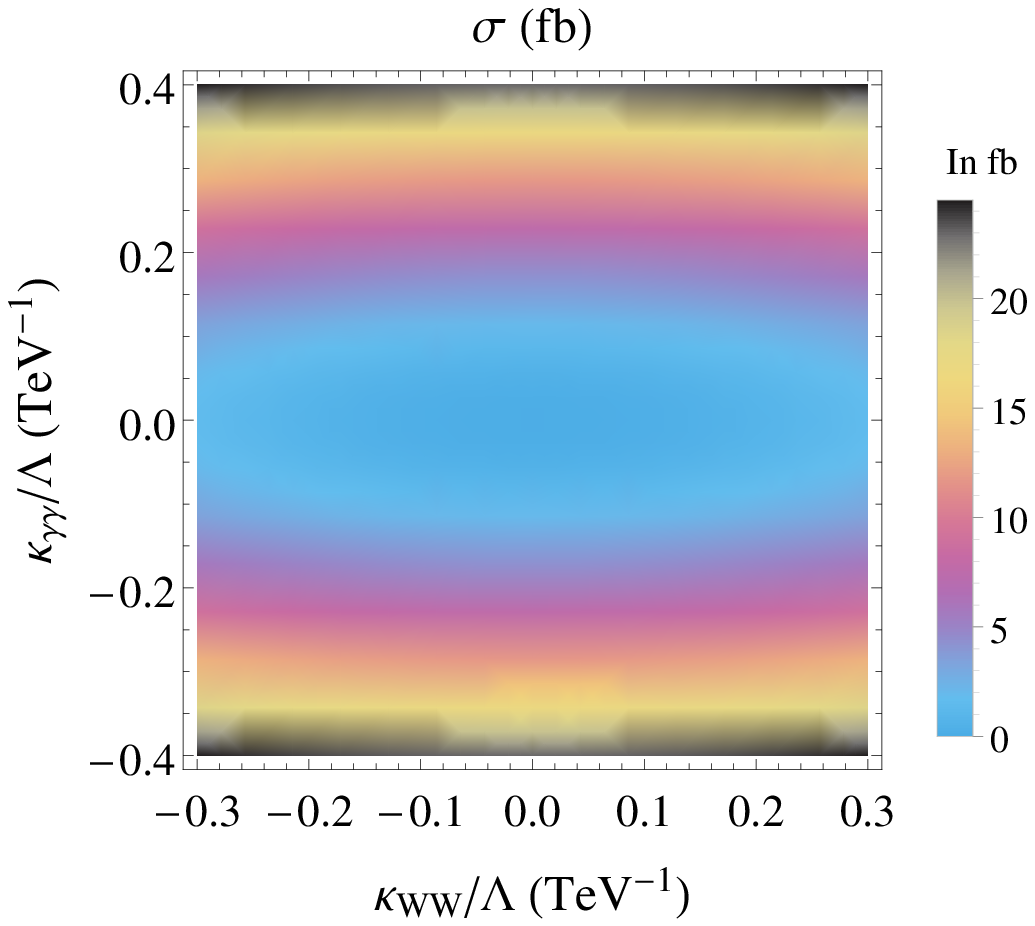}
\includegraphics[width=0.3\textwidth]{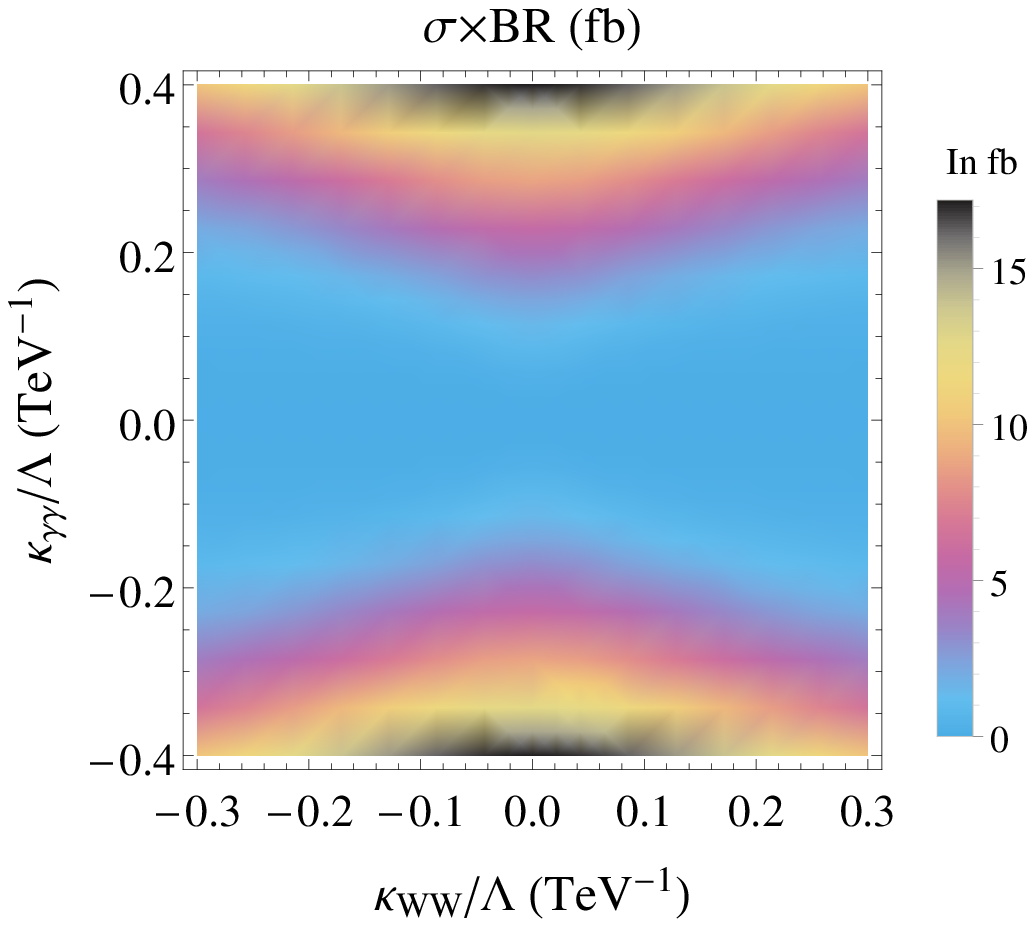}
\includegraphics[width=0.3\textwidth]{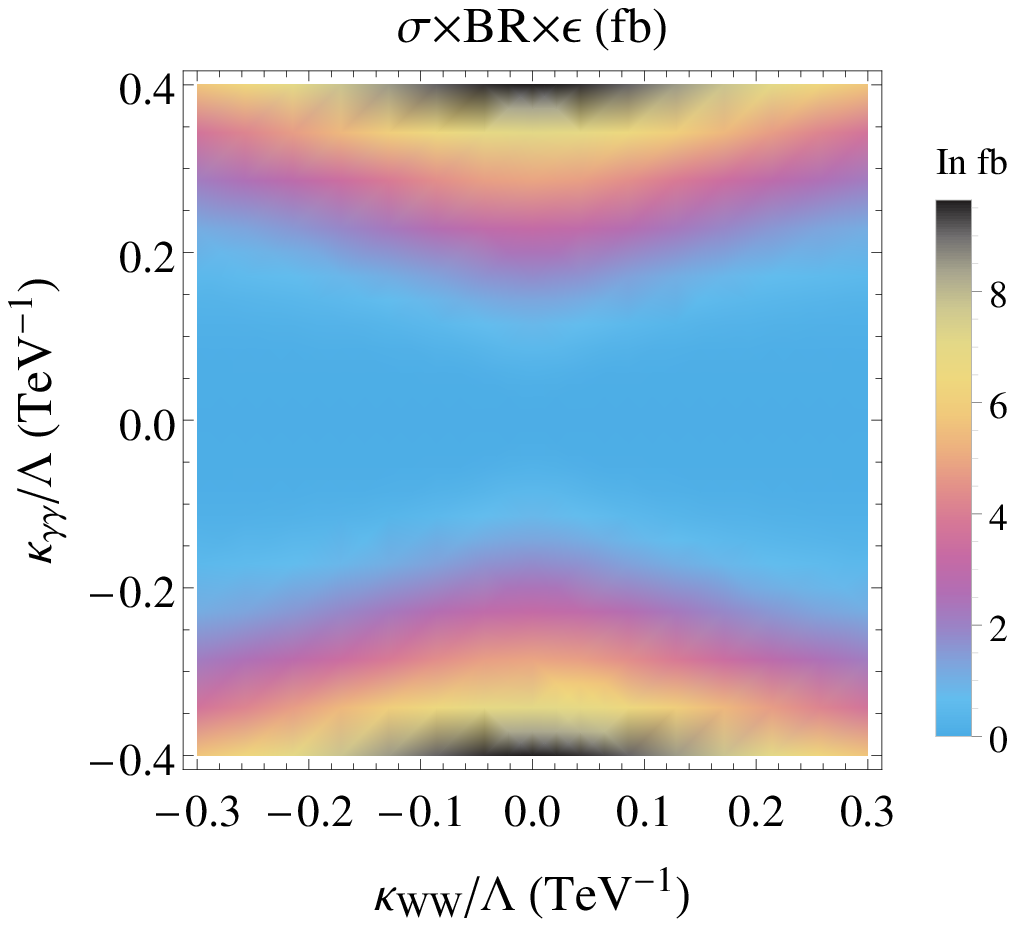}
\caption{2D plots of $\sigma$, $\sigma\times\mathrm{BR}$ and 
$\sigma\times\mathrm{BR}\times\ep$ for $M_{\Phi}=1$ TeV for two coupling assumptions - $\kp_{gg},\kp_{\gm\gm}\neq 0$ (first row) and 
$\kp_{WW},\kp_{\gm\gm}\neq 0$ (second row) while setting other $\kp_{xy}$ to zero.}
\label{fig:2D}
\end{figure}

In general, when $\kp_{\gm\gm}$ and any one $\kp_{xy}$ are nonzero, the production cross section
can be expressed as 
\be 
\label{eq:cs}
\sg\lt(M_{\Phi},\kp_{xy},\kp_{\gm\gm},\Lm\rt) = 
\Lm^{-2}\lt\{\kp_{xy}^2\sg^0_{xy}(M_{\Phi})
+\kp_{\gm\gm}^2\sg^0_{\gm\gm}(M_{\Phi}) 
+\kp_{xy}\kp_{\gm\gm}\sg^0_I(M_{\Phi})\rt\}\ ,
\ee
where the $\sg^0_I$ is the interference contribution and $\sg^0_{xy}$ ($\sg^0_{\gm\gm}$) in the \emph{r.h.s.} is the production cross sections 
through the $xy$ ($\gm\gm$) fusion (see Fig.~\ref{fig:cs}). These cross sections as functions of mass have been computed
numerically by interpolating cross sections points in the mass range $M_{\Phi}=0.5-2.5$ TeV. 
When more than one $\kp_{xy}$ are nonzero, the combined selection cut efficiency, in
general, depend on $M_{\Phi}$ and $\kp_{xy}$. Including cut efficiencies in 
Eq.~\eqref{eq:cs} (omitting the functional dependence on $M_{\Phi}$), we get
\be 
\label{eq:cseff}
\sg\times\ep = 
\Lm^{-2}\lt\{\kp_{xy}^2\sg^0_{xy}\ep_{xy}
+\kp_{\gm\gm}^2\sg^0_{\gm\gm}\ep_{\gm\gm}
+\kp_{xy}\kp_{\gm\gm}\sg^0_I\ep_I\rt\}\ ,
\ee
where $\ep_{xy}$, $\ep_{\gm\gm}$ are the cut efficiencies for the pure $xy$ and pure $\gm\gm$ fusion
production modes respectively and they are functions of $M_{\Phi}$ only. Whereas the combined efficiency 
$\ep$ and, $\ep_I$ associated with
the interference term are functions of $M_{\Phi}$, $\kp_{xy}$ and $\kp_{\gm\gm}$. 
We have seen
that $\ep_I$ is mostly sensitive to $M_{\Phi}$ but not to the couplings. Therefore, for simplicity
we use $\ep_I=\ep_{xy}(M_{\Phi})$ for $\kp_{xy}^2\sg^0_{xy} > \kp_{\gm\gm}^2\sg^0_{\gm\gm}$ region and 
$\ep_I=\ep_{\gm\gm}(M_{\Phi})$ for $\kp_{\gm\gm}^2\sg^0_{\gm\gm} > \kp_{xy}^2\sg^0_{xy}$ region.
Branching fraction in the
$\gm\gm$ channel can be expressed as
\be 
\label{eq:BRyy}
\mathrm{BR}_{\gm\gm}\lt(M_{\Phi},\kp_{xy},\kp_{\gm\gm}\rt) = 
\frac{\kp_{\gm\gm}^2\Gm_{\gm\gm}(M_{\Phi})}{\kp_{\gm\gm}^2\Gm_{\gm\gm}(M_{\Phi}) + \kp_{xy}^2\Gm_{xy}(M_{\Phi})}\ ,
\ee
where $\Gm$'s are known analytically from Eq.~\eqref{eq:pwformula}. Finally, we derive exclusion regions in Fig.~\ref{fig:exclusion} by using Eqs.~\eqref{eq:cseff} and \eqref{eq:BRyy} in Eq.~\eqref{eq:Ns}.

In Fig.~\ref{fig:cecom}, we show combined cut efficiency for the coupling assumption $\kp_{gg},\kp_{\gm\gm}\neq 0$
(all other $\kp_{xy}$ are zero). This combined cut efficiency should lie between the two individual efficiencies 
$\ep_{\gm\gm}\sim 55\%$ and $\ep_{gg}\sim 60\%$ according to its definition. 
One might also be interested to see the behavior of $\sigma$, $\sigma\times\mathrm{BR}$ and 
$\sigma\times\mathrm{BR}\times\ep$ for different coupling assumptions. In Fig.~\ref{fig:2D}, we
show these three quantities in the 2D plane for the two cases - $\kp_{gg},\kp_{\gm\gm}\neq 0$ and 
$\kp_{WW},\kp_{\gm\gm}\neq 0$ (other $\kp_{xy}$ are set to zero).

\subsection{Distinguishing different production modes}

\begin{figure}[h!]
\includegraphics[height=5.5cm,width=7cm]{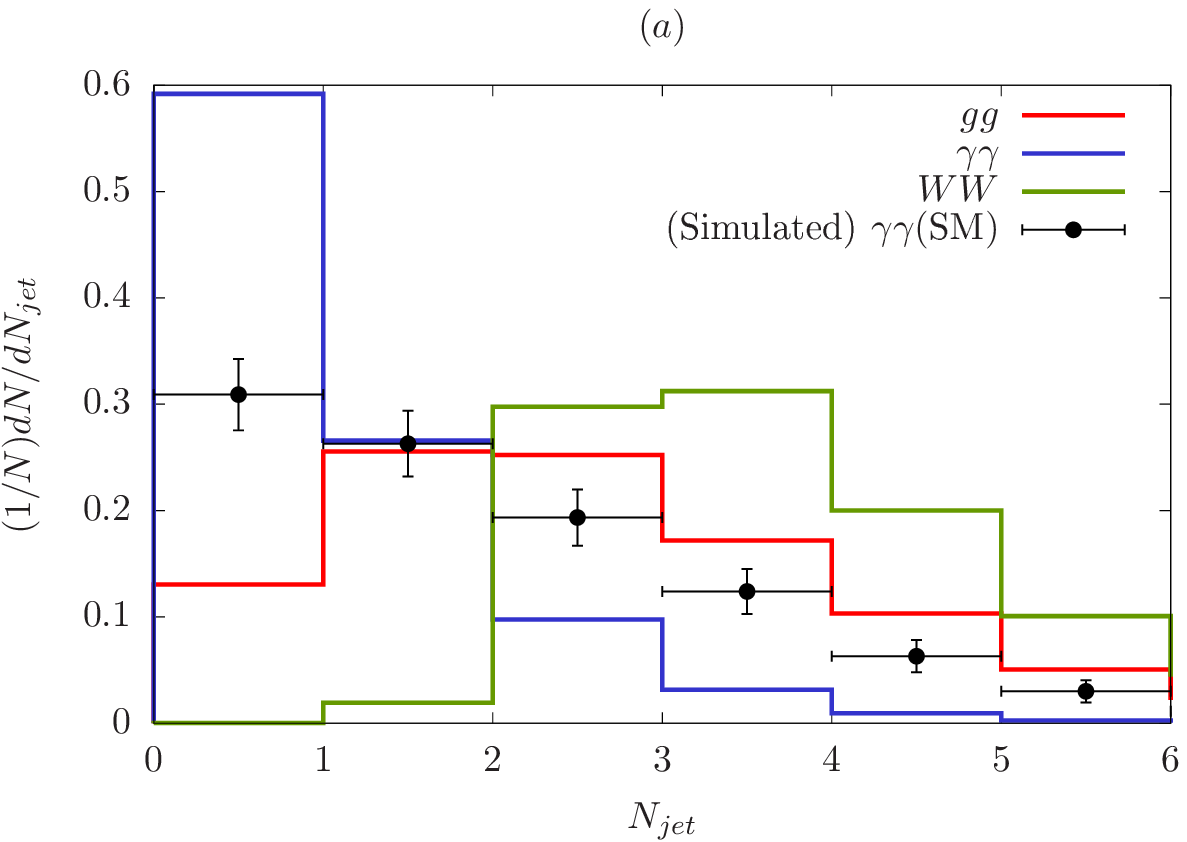}\hspace{0.25cm}
\includegraphics[height=5.5cm,width=7cm]{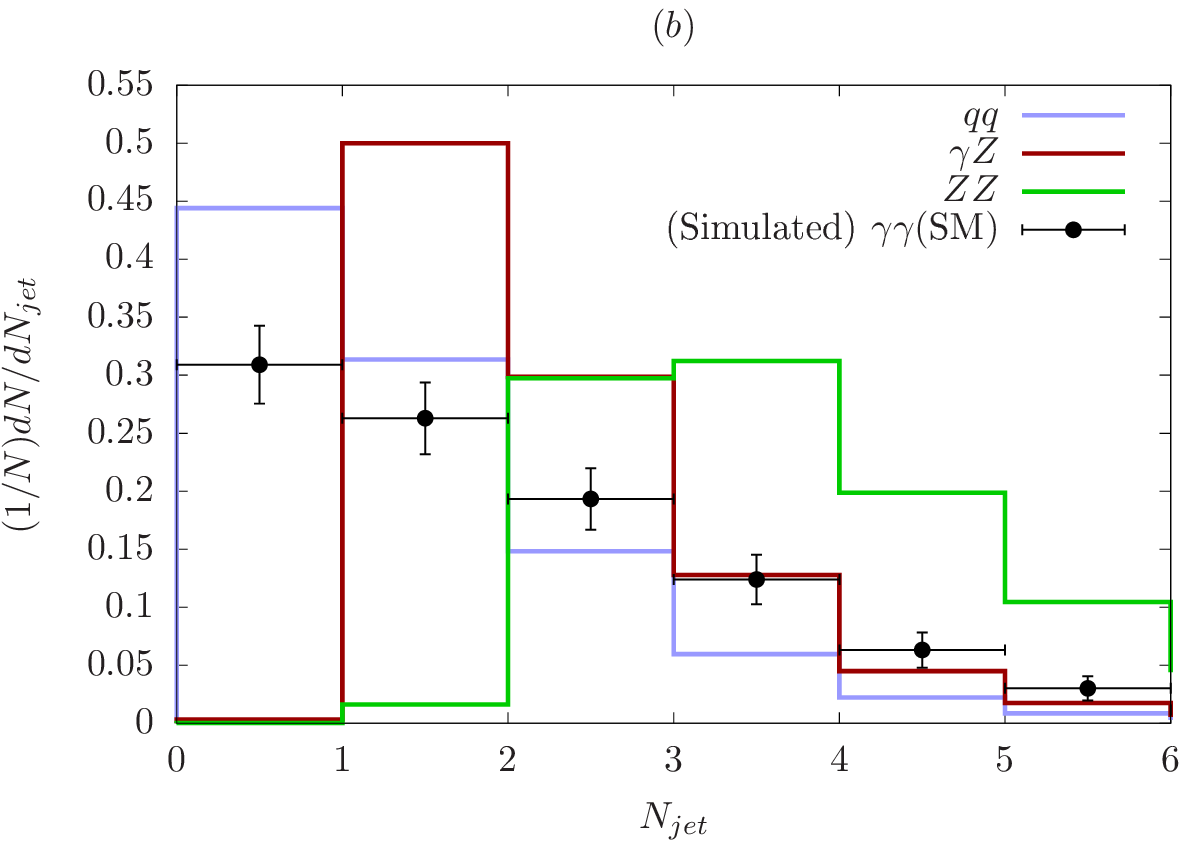}
\caption{Jet multiplicity ($N_{jet}$) distributions of various production modes of the scalar
$\Phi$ for $M_{\Phi}=1$ TeV at the 13 TeV LHC. The $N_{jet}$ distribution for the dominant
SM background {\it i.e.} for the $pp\to\gm\gm$ process is obtained from our simulation.
The uncertainty associated with this distribution is statistical uncertainty only which is 
computed for $\mc{L}=50$ fb$^{-1}$.}
\label{fig:NJND}
\end{figure}

A common way to distinguish different production modes of a heavy scalar is to scrutinize 
various kinematic distributions especially the jet activities associated with the scalar. 
It was pointed out in 
Refs.~\cite{Csaki:2016raa,Harland-Lang:2016qjy,Ebert:2016idf,Dalchenko:2016dfa,Fuks:2016qkf}
that the jet multiplicity ($N_{jet}$) distribution could be very important in this regard. 
In Fig.~\ref{fig:NJND}, we show the normalized $N_{jet}$ distributions
for various production modes of the scalar and compare them with the SM prediction. These distributions are obtained assuming $M_{\Phi}=1$ TeV
at the 13 TeV LHC with 50 fb$^{-1}$ integrated luminosity with 
the diphoton invariant mass ($M_{\gm\gm}$) satisfying $|M_{\gm\gm}-M_{\Phi}|< 150$ GeV,
in addition to the set of cuts defined earlier. Our jet selection cuts are $p_T(j)>25$ GeV for $|\eta(j)|<2.5$ and
$p_T(j)>50$ GeV for $|\eta(j)|>2.5$.
The dominant background 
contribution of about 90\% comes from the SM $q\bar{q}\to\gm\gm$
process and, in this analysis, we only consider this as the background which we estimate from our simulation. The error bars
associated with the background represent the statistical uncertainly only. In reality, various components of systematic
uncertainties like the jet energy scale, jet energy resolution, uncertainty in the luminosity must also be considered to obtain the total uncertainty~\cite{Khachatryan:2016yec}.
But the systematic uncertainty becomes small compared to the statistical one when 
background distributions are obtained from data. 

\begin{table}[h!]
\begin{tabular}{|c|c|c|c|c|c|c|}
\hline
$gg\to\Phi$ & $qq\to\Phi$ & $\gm\gm\to\Phi$ & $\gm Z\to\Phi$ & $WW\to\Phi$ & $ZZ\to\Phi$ & $\gm\gm$ SM\\
\hline
1.94 & 0.92 & 0.61 & 1.75 & 2.85 & 2.87 & 1.42\\
\hline
\end{tabular}
\caption{Average jet multiplicity for different production modes of $\Phi$ at the 13 TeV LHC with
$\mc{L}=50$ fb$^{-1}$. Average jet multiplicity is defined by the sum, $\sum_i (BH)_iN_i$ where $BH_i$ represents
the bin height of the $i$-th bin of the normalized $N_{jet}$ distribution and $N_i=i-1$ is the number of jets associated
with the $i$-th bin.}
\label{tab:avgnj}
\end{table}

It is visibly 
clear that the different production modes display very different jet multiplicity distributions.
The distributions for the $\gm\gm$ and the $qq$ fusion modes peak at 0-jet but the peak for the $qq$ mode is
not as sharp as the $\gm\gm$ mode. Cross section for the 0-jet bin
for the $\gm\gm$ mode is roughly about 60\% of the total cross section. On the other hand, it is about
45\% for the $qq$ fusion case. The SM background $N_{jet}$ distribution also peaks at 0-jet, but contains 
only 30\% of the total cross section.
The $gg$ fusion shows a peak 
at 1 and 2-jet whereas the vector boson fusion production through the $WW$ and $ZZ$ fusions show peak at 2 and 3-jet.
The $\gm Z$ fusion mode, on the other hand, shows a peak at 1-jet. 

Different nature of the $N_{jet}$ distribution can be captured by the average jet multiplicity associated with the 
diphoton resonance. We compute the average jet multiplicity of different production modes and the background, and 
report these numbers in Table~\ref{tab:avgnj}. It is expected that if the scalar is produced
through the $\gm\gm$ fusion, then the average jet multiplicity is lower compared to the
$gg$ or the $qq$ fusions. This is because, in case for the $\gm\gm$ fusion, a hard jet in the final state
can originate from the $q\to q\gm$ splitting. However, this is suppressed compared to the 
leading order (LO) process with zero jet by the small probability of $q\to q\gm$
branching and also by the smallness of $\al$. On the other hand, colored particles in the initial state, {\it i.e.} 
in case for the $gg$ or the $qq$ fusions, leads to higher jet multiplicity.
The average jet multiplicity is greater than two for the
Vector boson fusion modes because one would anticipate to get
at least two hard jets in most of the events since two initial $V$'s come from the $q\to q'V$ branching.
For the
$\gm Z$ initial state, one expects at least one hard jet from the $q\to qZ$ splitting.

\subsection{Multivariate analysis}

\begin{figure}[h!]
\includegraphics[width=5cm,height=4cm]{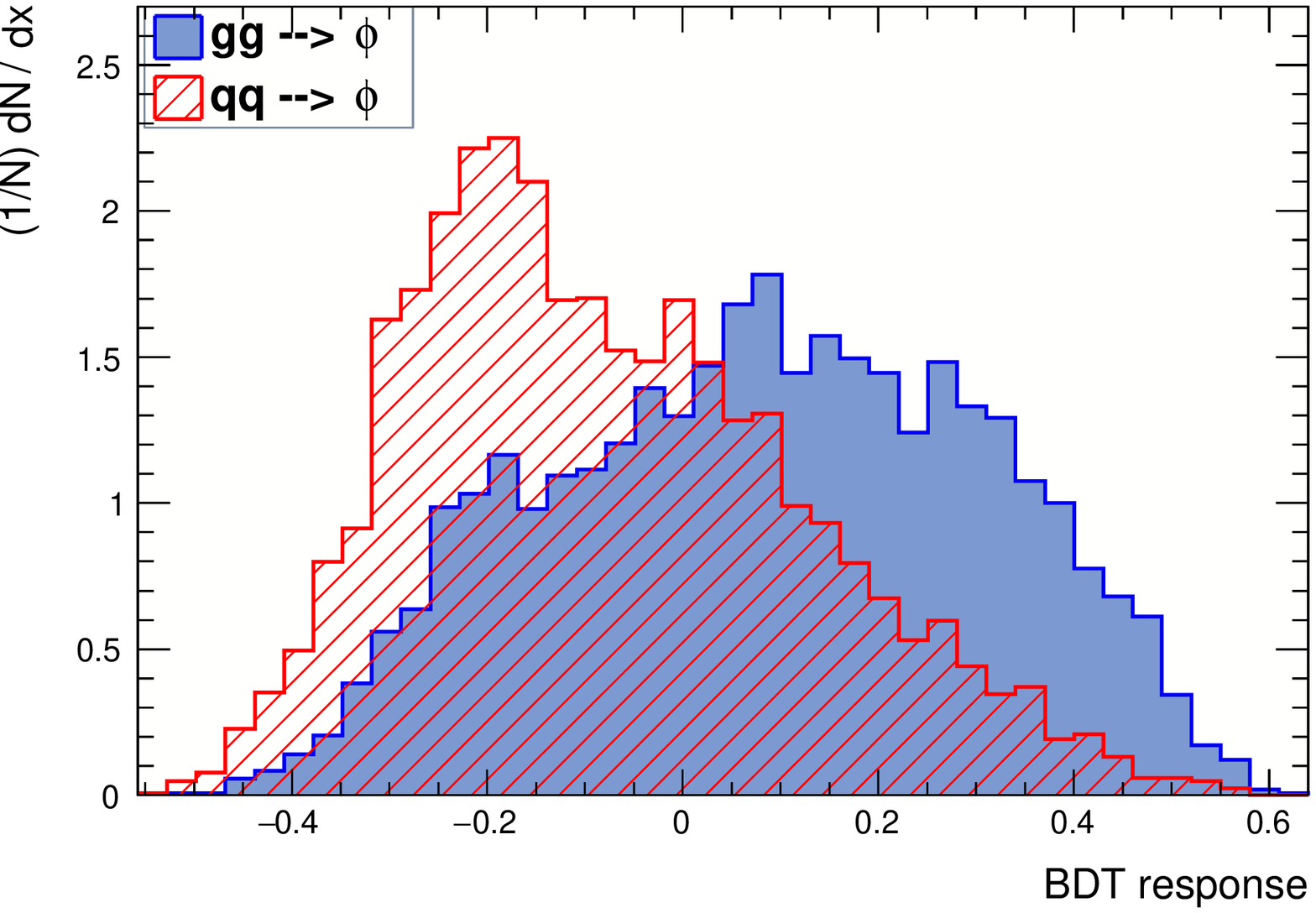}
\includegraphics[width=5cm,height=4cm]{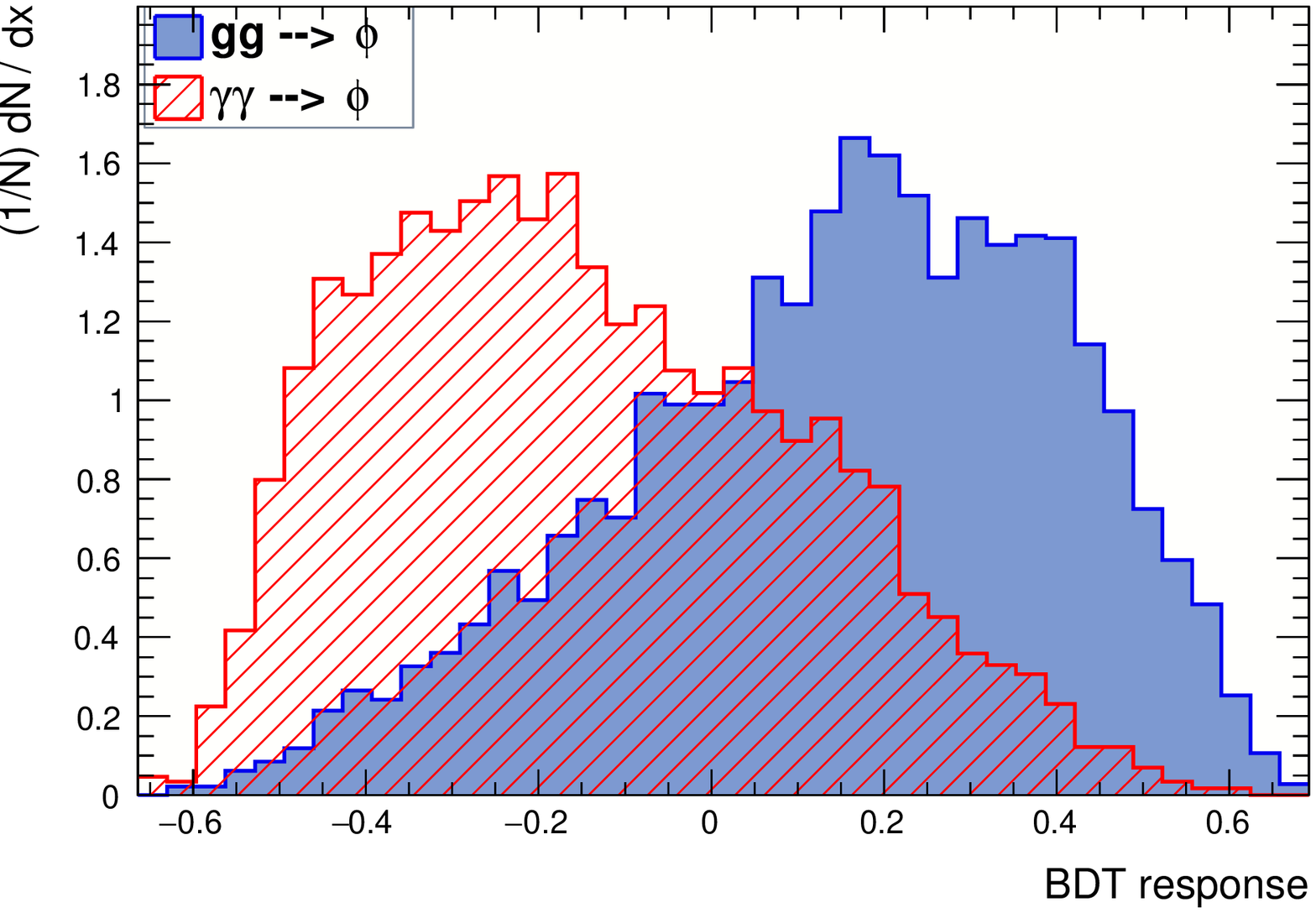}
\includegraphics[width=5cm,height=4cm]{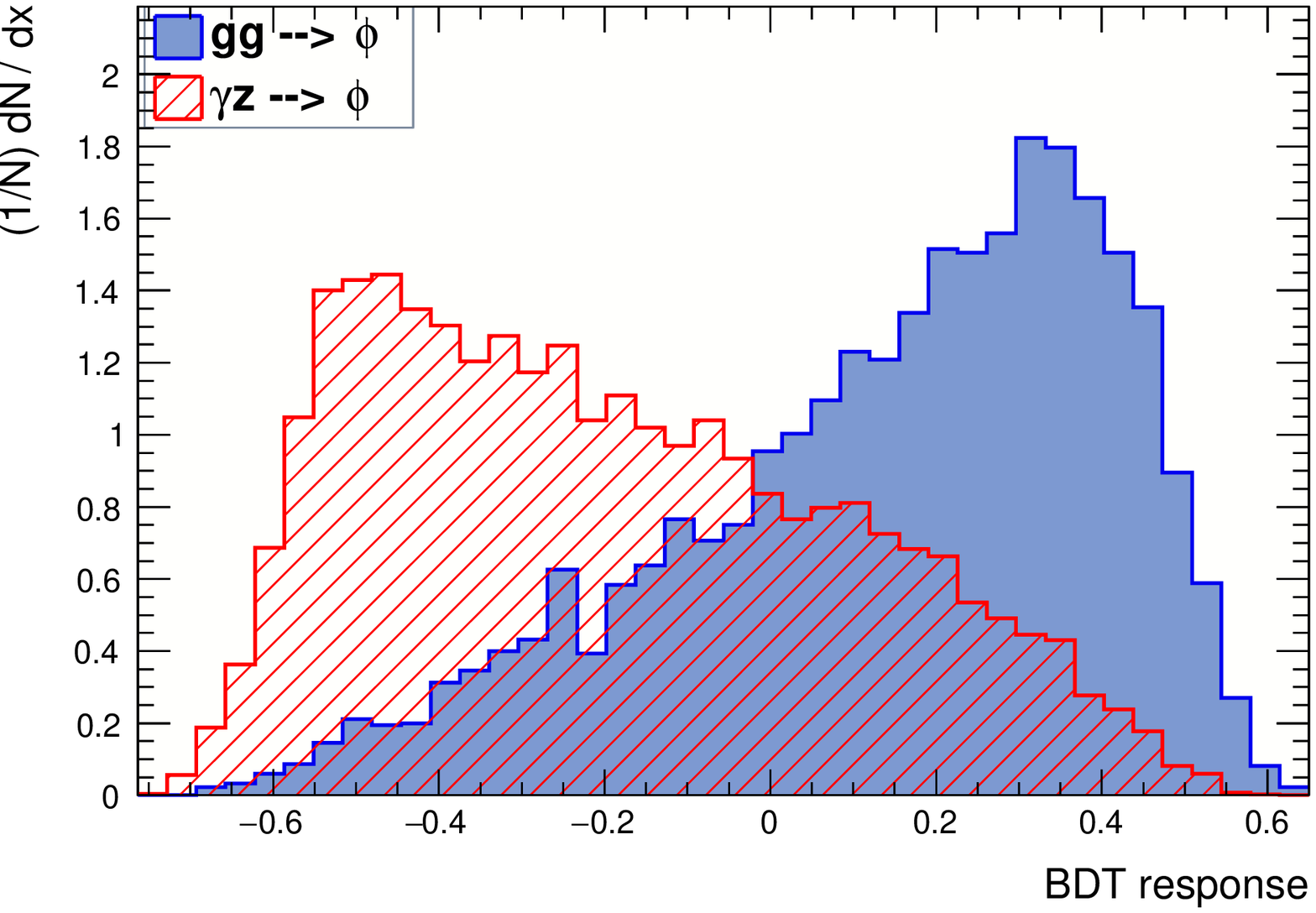}\\
\includegraphics[width=5cm,height=4cm]{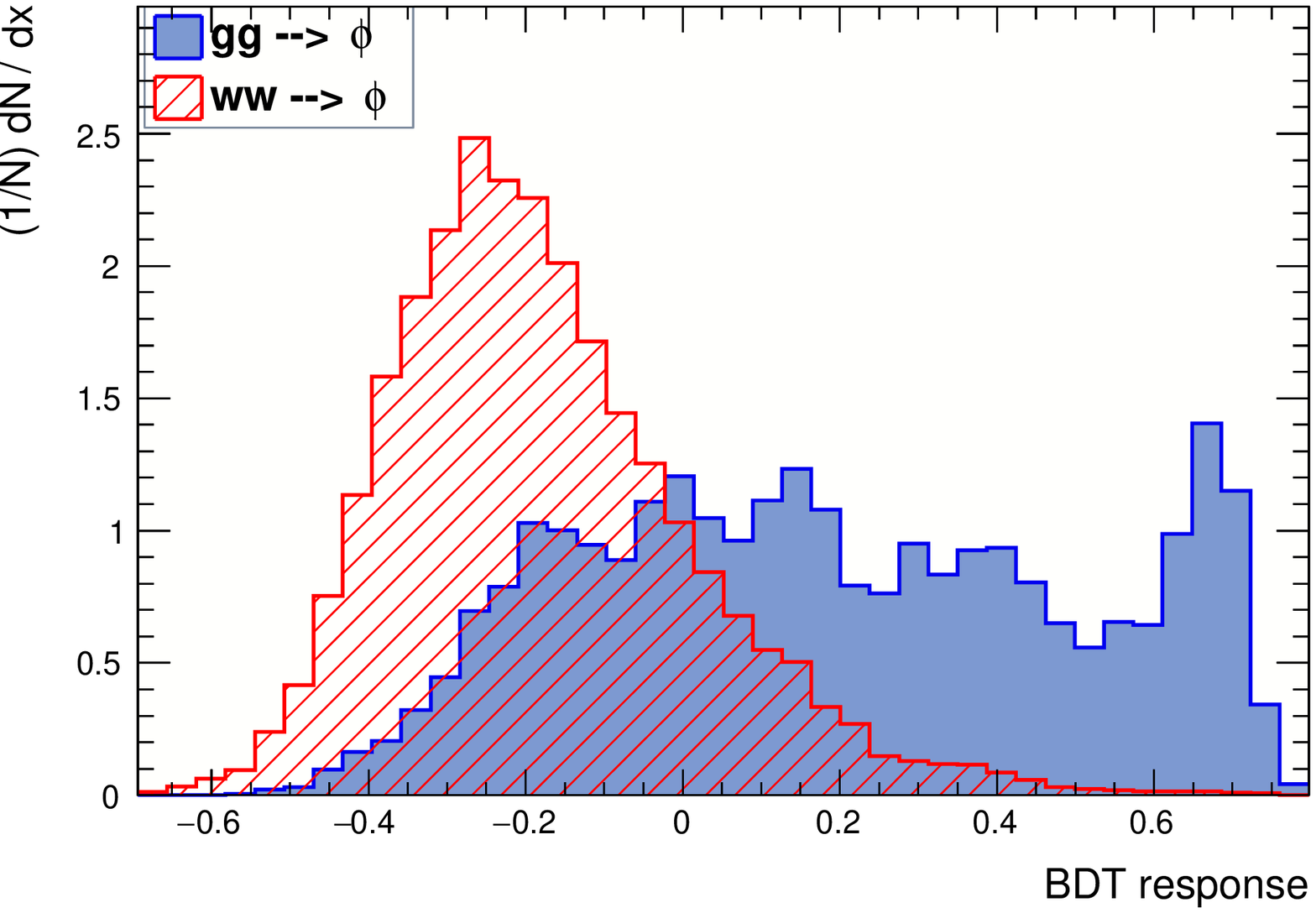}
\includegraphics[width=5cm,height=4cm]{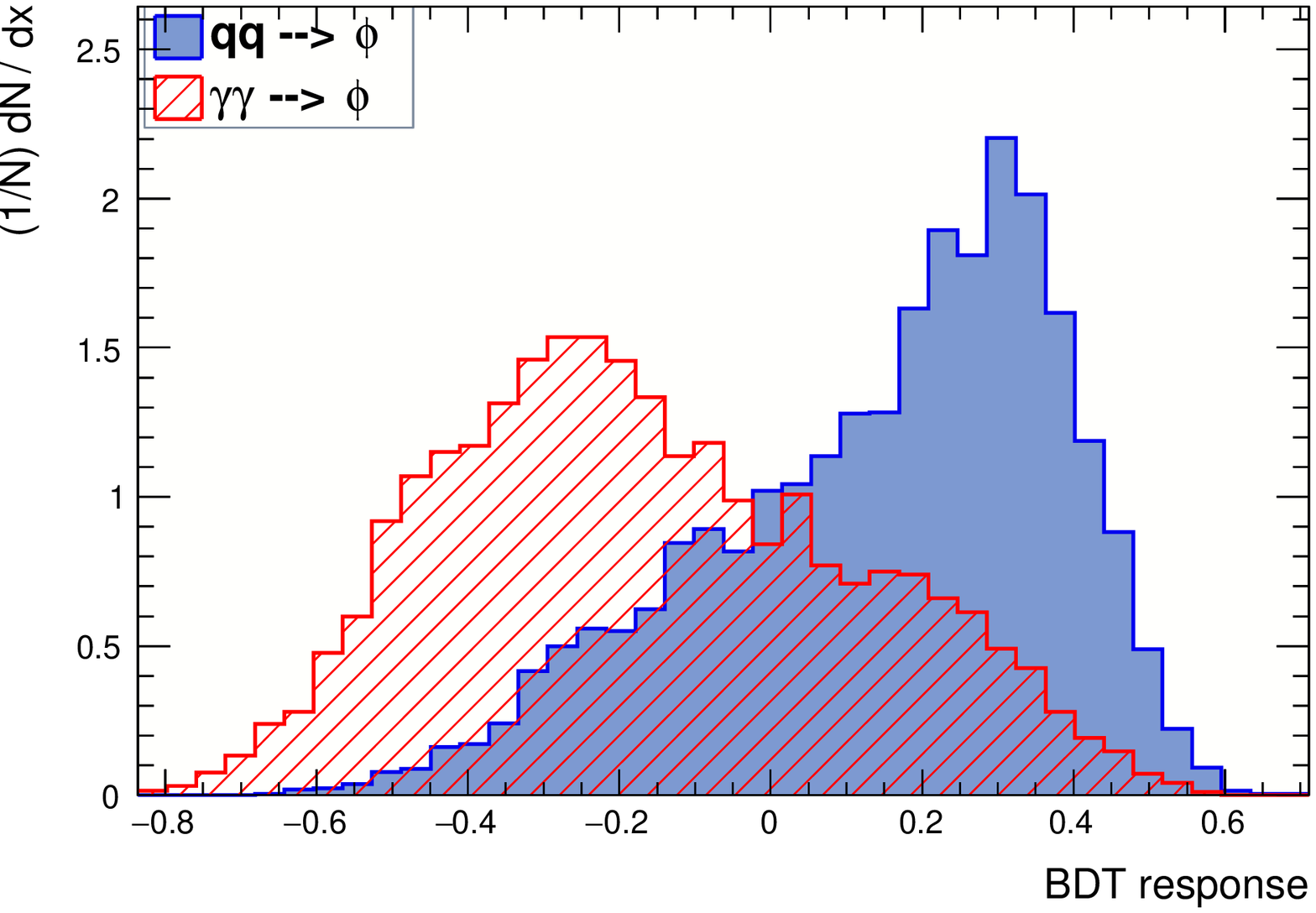}
\includegraphics[width=5cm,height=4cm]{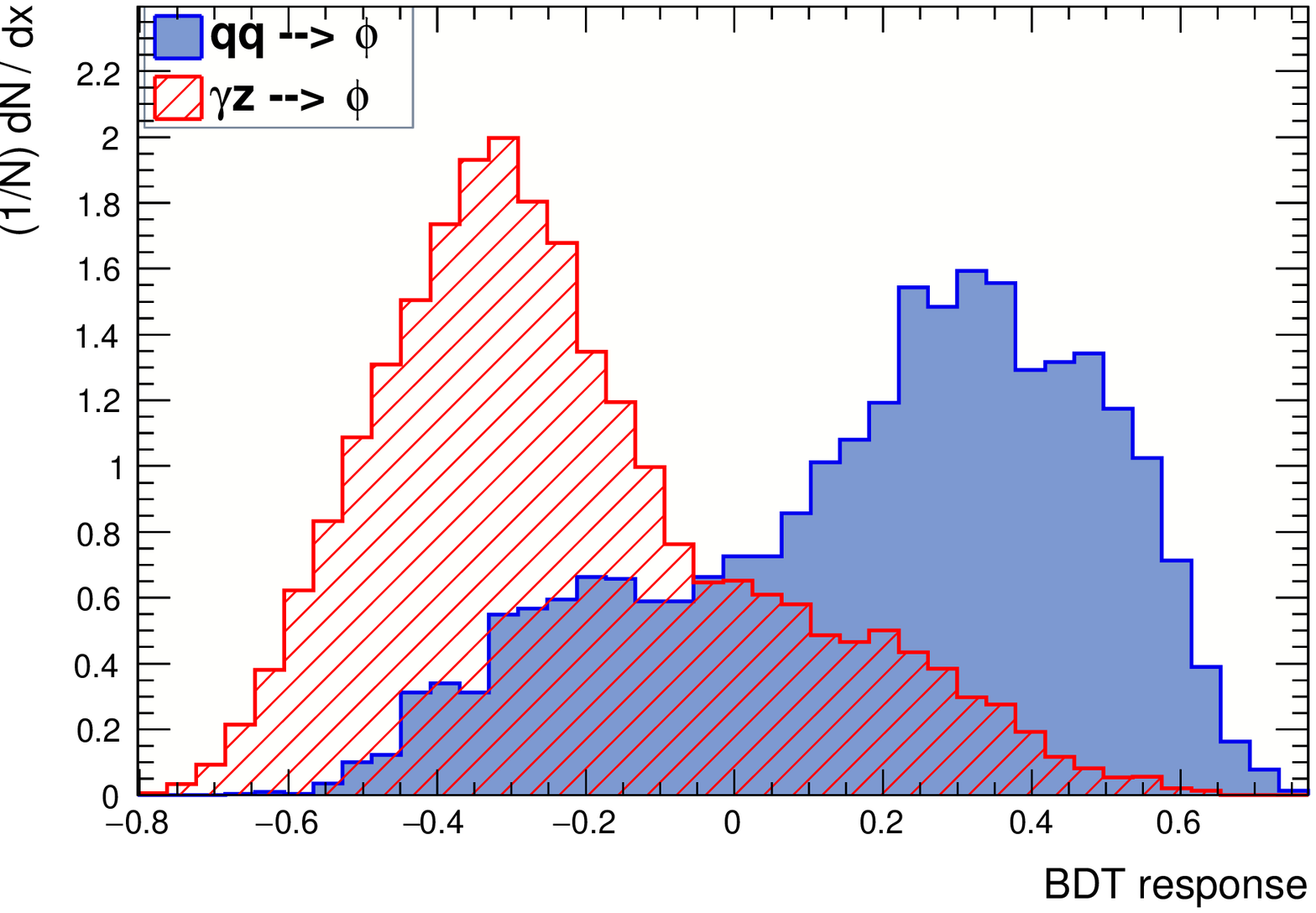}\\
\includegraphics[width=5cm,height=4cm]{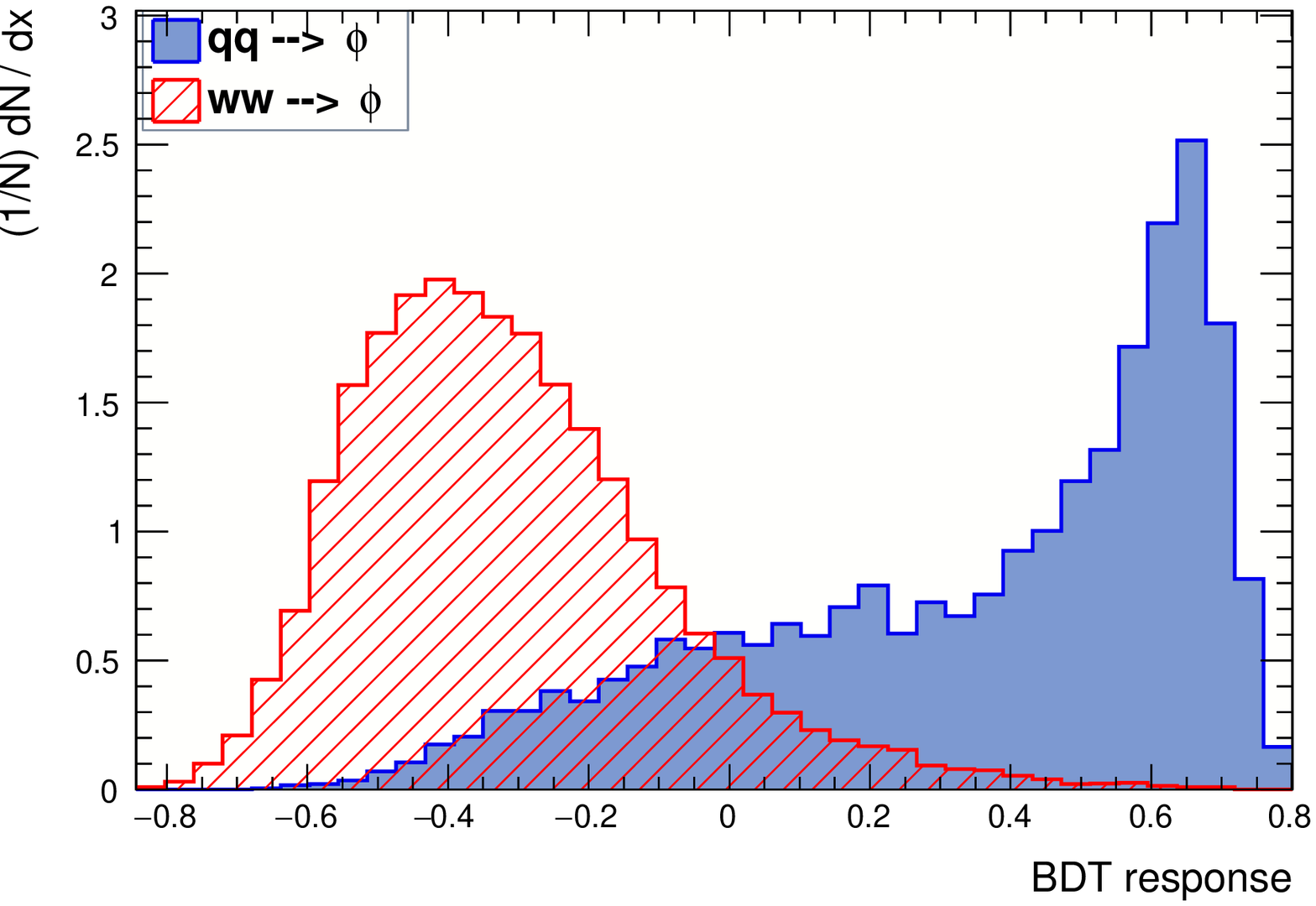}
\includegraphics[width=5cm,height=4cm]{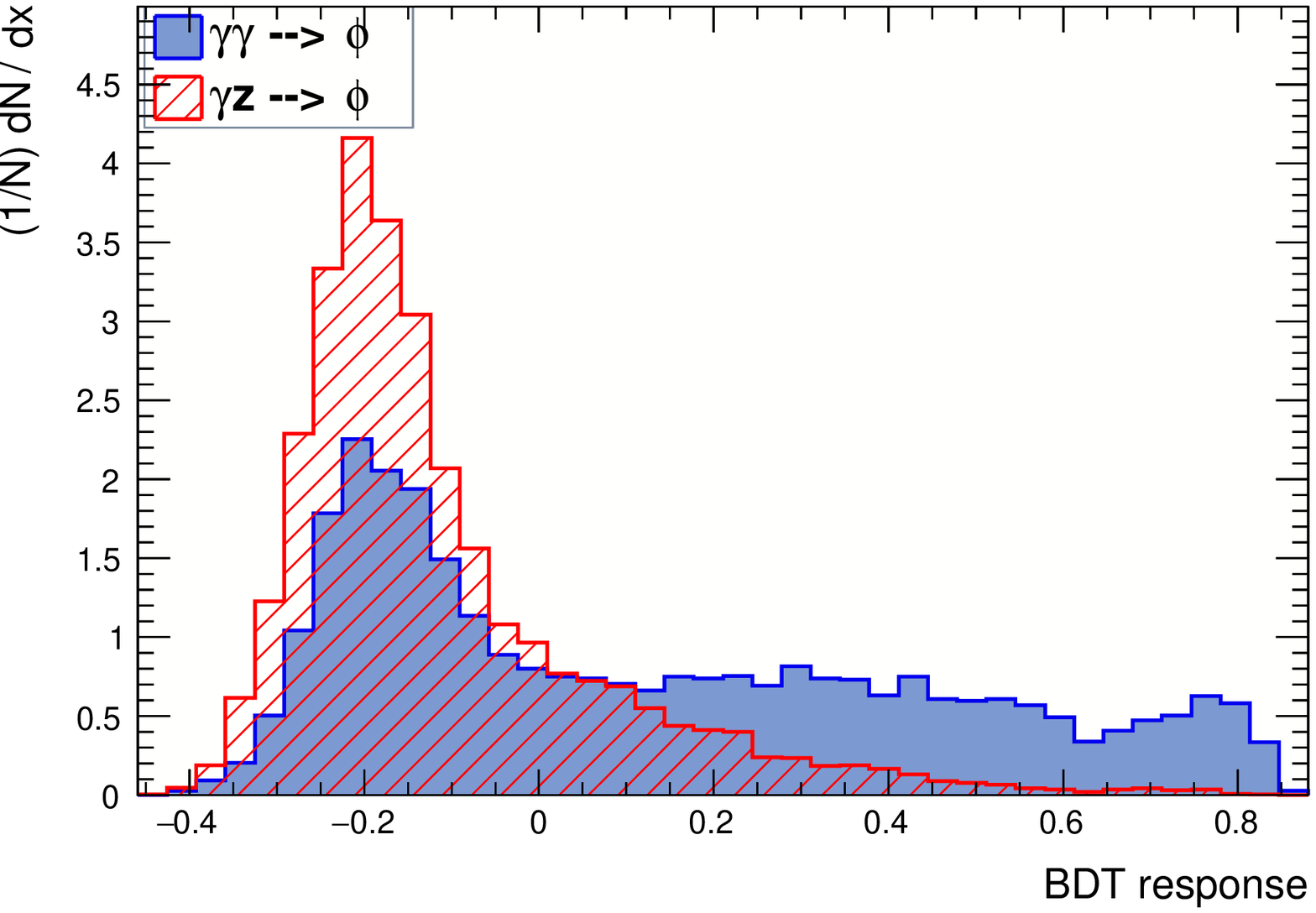}
\includegraphics[width=5cm,height=4cm]{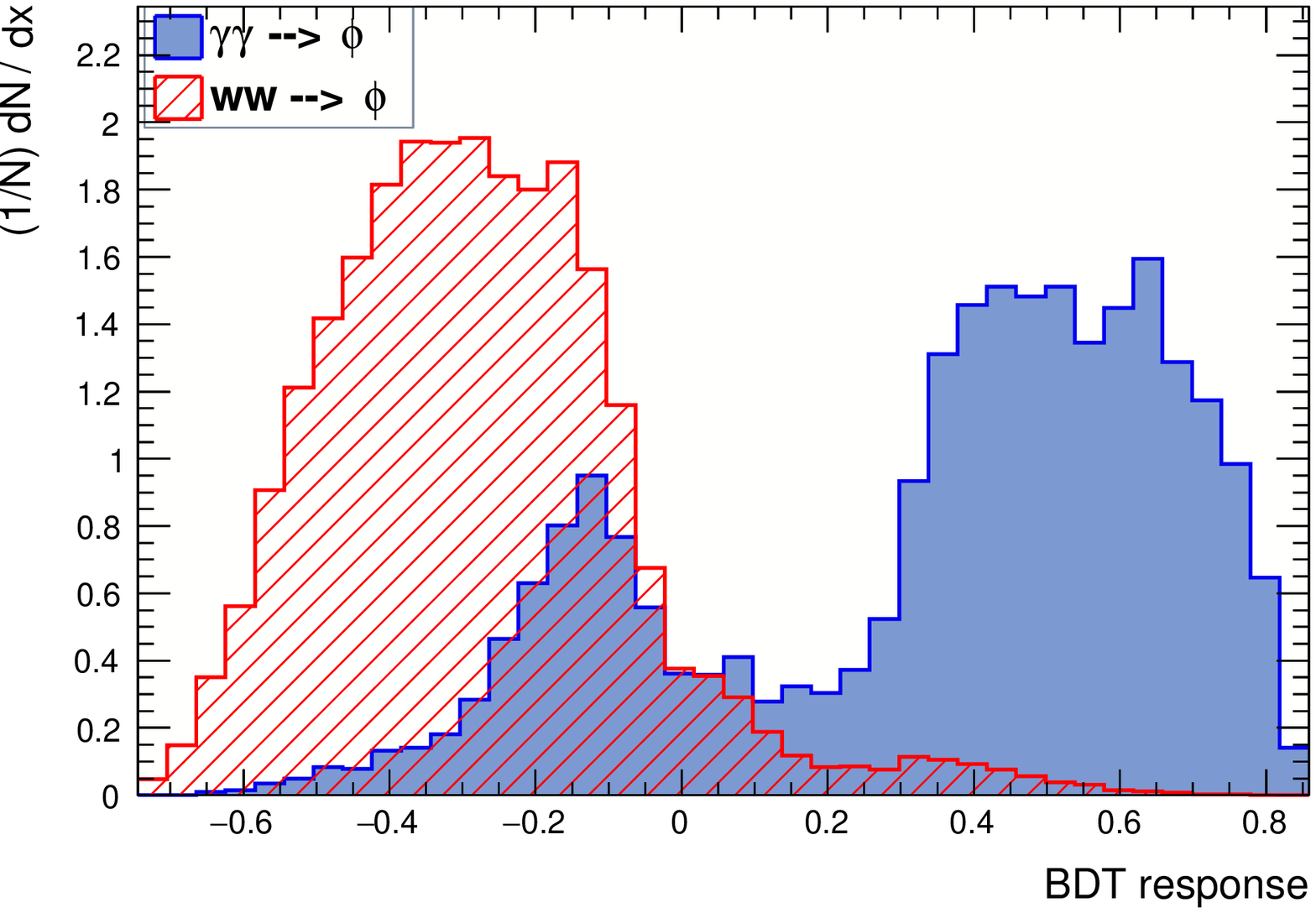}\\
\includegraphics[width=5cm,height=4cm]{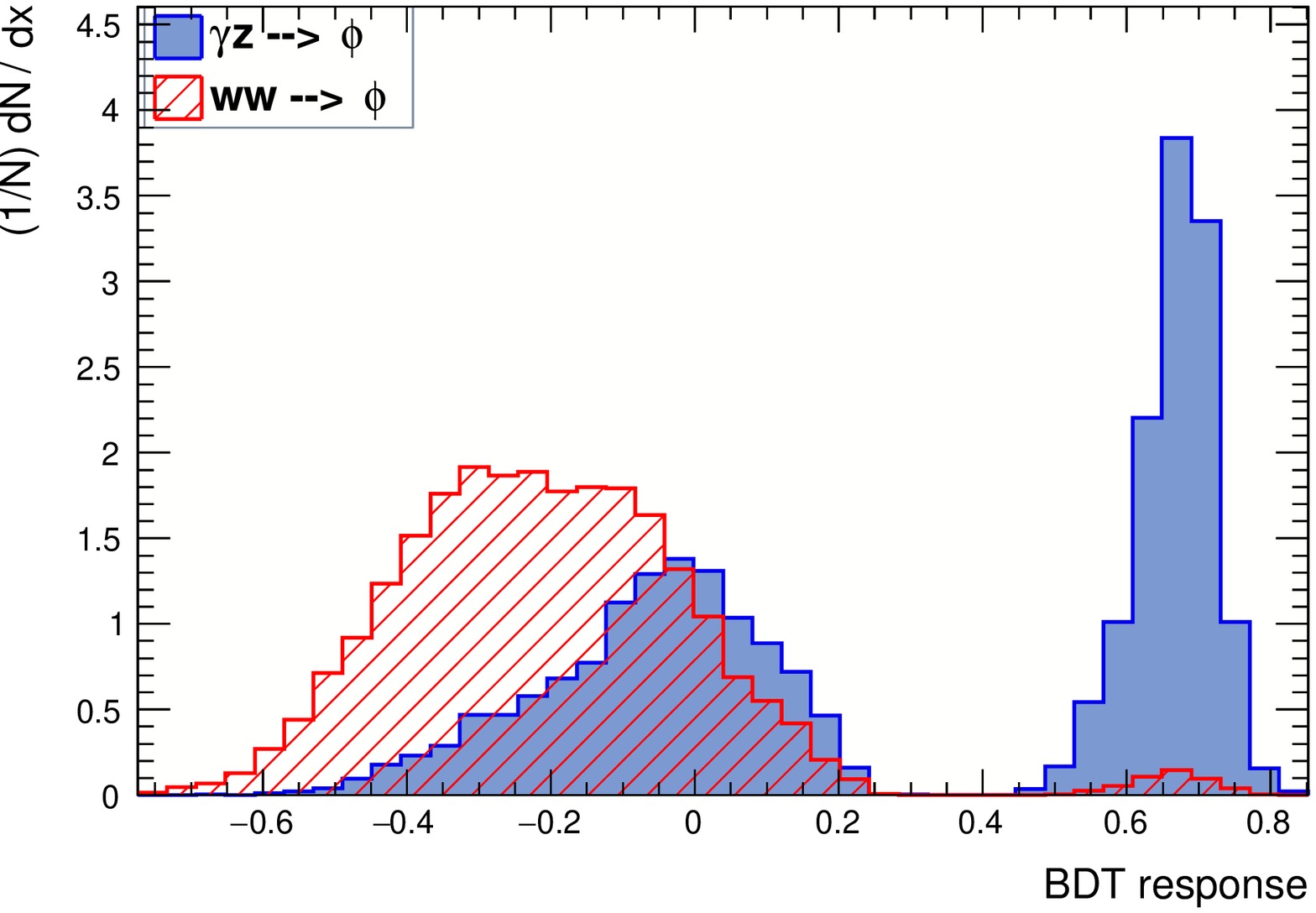}
\caption{BDT responses of two different production modes at a time. These are obtained using a MVA with
twelve input variables as listed in Table~\ref{tab:MVAvar} for $M_{\Phi}=1$ TeV at the 13 TeV LHC.
Diphoton events are selected by applying $|M_{\gm\gm}-M_{\Phi}|< 150$ GeV cut in addition to the ATLAS 
selection cuts as listed in the text.}
\label{fig:BDT}
\end{figure}

In the previous subsection, we show as a demonstration that the jet multiplicity distributions of two different 
production modes (and also for the background) can be quite different. 
Apart from the $N_{jet}$ distribution, there are other kinematic
variables which also show some differences in their shapes for different production modes.
For example, in Ref.~\cite{Csaki:2016raa}, the authors showed that various distributions like 
the scalar sum of transverse
energy $H_T$, pseudorapidity ($\eta$) of the selected photons and jets, central rapidity gap ($\Dl\eta$)
between the jets and the scalar show some visible differences for 
the $gg$ and the $\gm\gm$ production modes.

\begin{table}[!h]
\centering
\begin{tabular}{|c|c|c|c|c|c|c|c|}
\hline 
Variable & Importance & Variable & Importance & Variable & Importance & Variable & Importance \\ 
\hline 
$N_{jet}$    & $1.8 \times 10^{-1}$ & $p_T(\gm_1)$ & $4.9 \times 10^{-2}$ & $\eta(\gm_1)$ & $7.5 \times 10^{-2}$ & $\Dl R(\gm_1,\gm_2)$ & $6.7 \times 10^{-2}$ \\ 
\hline 
$H_T$ & $4.5 \times 10^{-2}$ & $p_T(\gm_2)$ & $6.0 \times 10^{-2}$ & $\eta(\gm_2)$ & $6.2 \times 10^{-2}$ & $\Dl R(\gm_1,j_1)$ & $9.9 \times 10^{-2}$ \\ 
\hline 
$\Dl\eta(\Phi,j_1)$  & $8.1 \times 10^{-2}$ & $p_T(j_1)$ & $9.1 \times 10^{-2}$ & $\eta(j_1)$ & $9.4 \times 10^{-2}$ & $\Dl R(\gm_2,j_1)$ & $1.0 \times 10^{-1}$ \\ 
\hline 
\end{tabular} 
\caption{Input variables used for MVA to separate $gg$ and $\gm\gm$ production modes and their relative importance.}
\label{tab:MVAvar}
\end{table}

A cut based analysis which employs a set of rectangular cuts may not perform well to decipher the 
underlying production mechanism of the scalar.
In order to effectively distinguish two different production modes, one can use various kinematic variables 
that show some (small) differences in their shapes simultaneously in a MVA
whose output might show large differences in their shapes. If appropriate variables are chosen,
a MVA is expected to perform better than a cut-based analysis. Generally, MVA techniques are used 
to separate signal from background. Here, we use a MVA technique (BDT) to distinguish two different
production mechanisms more efficiently than a simple cut-based analysis. 
In particular, we use the adaptive BDT algorithm in the TMVA framework. 
We train the algorithm by tuning various parameters like the number of trees, minimum size of the node etc. 
for proper training of different production modes. Optimal values of these parameters are not fixed and they can 
differ for each analysis.

For MVA, we select events with at least one jet and construct twelve simple kinematic variables
as shown in Table~\ref{tab:MVAvar}. This includes $N_{jet}$, $H_T$, $\Dl\eta$ between 
$\Phi$ and leading jet, $\eta$ and $p_T$ of two selected photons and the leading jet and the separation $\Dl R$
in the $\eta-\phi$ plane between the photons and the leading jet. These twelve variables are finalized from a 
bigger set of variables by looking at their discriminatory power and less-correlation. 
In particular, the variables we use are not correlated more than $\sim 40\%$ for signal. But these
correlations might be different for the background.
Next to each variable in Table~\ref{tab:MVAvar}, we show their
relative importance in the BDT response and these numbers are obtained from TMVA using the $gg$ and the $\gm\gm$
production modes. Relative importance is a fraction (with all importance sums up to unity) which is used to
identify the ranking of the variables in MVA. In other words, greater relative importance of a variable signifies 
that the variable is a better discriminator. For actual definition of relative importance, interested readers may
look into the TMVA manual.
From Table~\ref{tab:MVAvar}, we see that $N_{jet}$ is the best discriminator to differentiate 
the $gg$ and the $\gm\gm$ production modes. Other variables like $\Dl R(\gm_1,j_1)$, $\Dl R(\gm_2,j_1)$,
$p_T(j_1)$, $\eta(j_1)$ and $\Dl\eta(\Phi,j_1)$ also act as good discriminators.
Here, our main aim is to distinguish different production mechanisms of the scalar using a suitable
MVA. Before arriving to this step, one might be interested to
see comparisons of various kinematic distributions for the different signal modes with the background.
In Appendix~\ref{app:dis}, we show distributions of some input variables for the signal and the background 
for the interested readers.

It should be remembered that relative importance or in other words the ranking of a variable might change for different production modes and also for 
different parameters like $M_{\Phi}$, $\sqrt{s}$ etc. which can change the shape of the kinematic distributions.
It is important to mention that this set of twelve
variables used here may not be the optimal one. One can always improve the analysis
with cleverer choices of variables.  

In Fig.~\ref{fig:BDT}, we show the BDT response by comparing two different production modes at a time.
The $WW$ and the $ZZ$ fusion modes are very similar in nature and, therefore, it is extremely difficult
to distinguish them. We do not consider the $ZZ$ fusion further as it is very much identical to the $WW$ mode.
We show, by picking two production modes at a time, ten such possible BDT responses in 
Fig.~\ref{fig:BDT}. 
These responses are substantially different for most of the combinations and therefore can be
distinguished very efficiently. We observe that it is hard to distinguish the $\gm\gm$ and the $\gm Z$
production modes as their BDT responses are not very different
from each other. One should also notice that there are two peaks in the BDT response of the $WW$ mode. This is because two
types of different topologies {\it i.e.} the associated production and the vector boson fusion 
contribute to the $\Phi jj$ final state. In case for the bimodal distributions like this, one can use two different BDTs that are trained for two different topologies to further improve the analysis. This type of advanced analysis is beyond the scope of 
this paper.

As a side remark, one should always be careful about overtraining while using the BDT algorithm (or any other algorithm
which uses nonlinear cuts). This can
happen without the proper choices of the algorithm specific tuning parameters. One can check
whether a test sample is overtrained or not by using the Kolmogorov-Smirnov (KS) statistics. 
Generally, if KS probability lies within the range 0.1 to 0.9 guarantees that the test sample 
is not overtrained. For this purpose, one uses two statistically independent samples, one for 
training and the other for testing.

\section{Conclusions}
\label{sec:conclu}

Among the various resonance search channels at the LHC, the diphoton channel is particularly 
important as this channel provides a comparatively cleaner background.
Generally, the diphoton resonance searches at the LHC assume that the resonance is produced
from the $gg$ fusion. Apart from the $gg$ fusion production, many BSM theories predict TeV-scale scalars 
that decay to diphotons can dominantly be produced by other means namely through the quark-quark ($qq$) fusion or 
through the gauge boson fusions ($\gm\gm$, $\gm Z$, $WW$ and $ZZ$).
In this paper we consider an effective field theory of a heavy scalar that decays to diphotons. 
In this model independent approach, the scalar can be produced in all the possible types mentioned above. 
We derive the exclusion limits on the mass and the effective couplings 
of the scalar using the latest 13 TeV ATLAS diphoton resonance search data with $\mc{L} = 36.7$ fb$^{-1}$.
While deriving the limits, we consider, for simplicity, only one effective coupling other than the 
$\kp_{\gm\gm}$ (since we only focus on the diphoton final state) is nonzero. 
We have properly taken care of the modified cut efficiencies while recasting the limits
set by the ATLAS collaboration. We find that when the scalar is dominantly produced from 
the $\gm\gm$ fusion, the latest LHC diphoton resonance search data sets limit on the new physics 
scale $\Lm\gtrsim 20$ TeV for the coupling $\kp_{\gm\gm}\sim 1$ for $M_{\Phi}\sim 1$ TeV.

In future, if a scalar resonance is seen at the LHC in the diphoton channel, the immediate
important issue one has to investigate that how the scalar is produced. Some preliminary 
analyses have already been done in the context of the 750 GeV resonance where it is shown
that the jet multiplicity distributions can be very different for the different production modes. 
In this paper we revisit the issue and show that the average jet multiplicity and
the $N_{jet}$ distribution can act as good discriminators. For better discrimination, we
use a sophisticated multivariate analysis by combining twelve simple kinematic variables to
distinguish one production mechanism from the other. Our analysis shows that 
one can identify different production mechanisms very efficiently at the LHC.

\section*{Acknowledgments}
The author thank Valery A. Khoze, Lucian A. Harland-Lang, Rikard Enberg and Gunnar Ingelman for helpful discussions.
This work is supported by the Swedish Research Council under contracts 621-2011-5107 and 2015-04814 and by the Carl Trygger Foundation under contract CTS-14:206.

\appendix
\section{Distributions for signal and background}
\label{app:dis}

\begin{figure}[h!]
\includegraphics[scale=0.65]{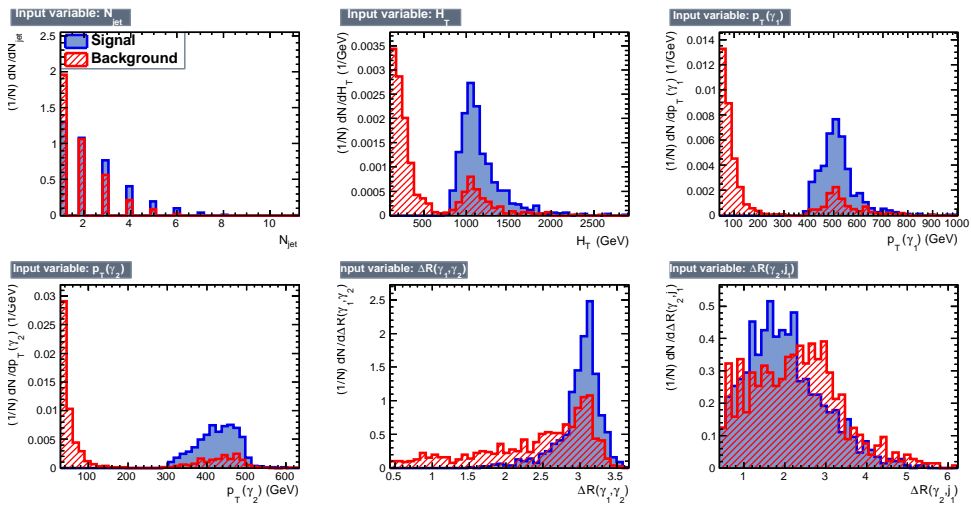}
\caption{Normalized distributions of a few sample input variables for the signal ($gg$ fusion with $M_{\Phi}=1$ TeV) and
background at the LHC ($\sqrt{s}=13$ TeV). These distributions are obtained by applying the selection cuts defined in Section~\ref{ssec:exclu}.}
\label{fig:vargg}
\end{figure}

\begin{figure}[h!]
\includegraphics[scale=0.65]{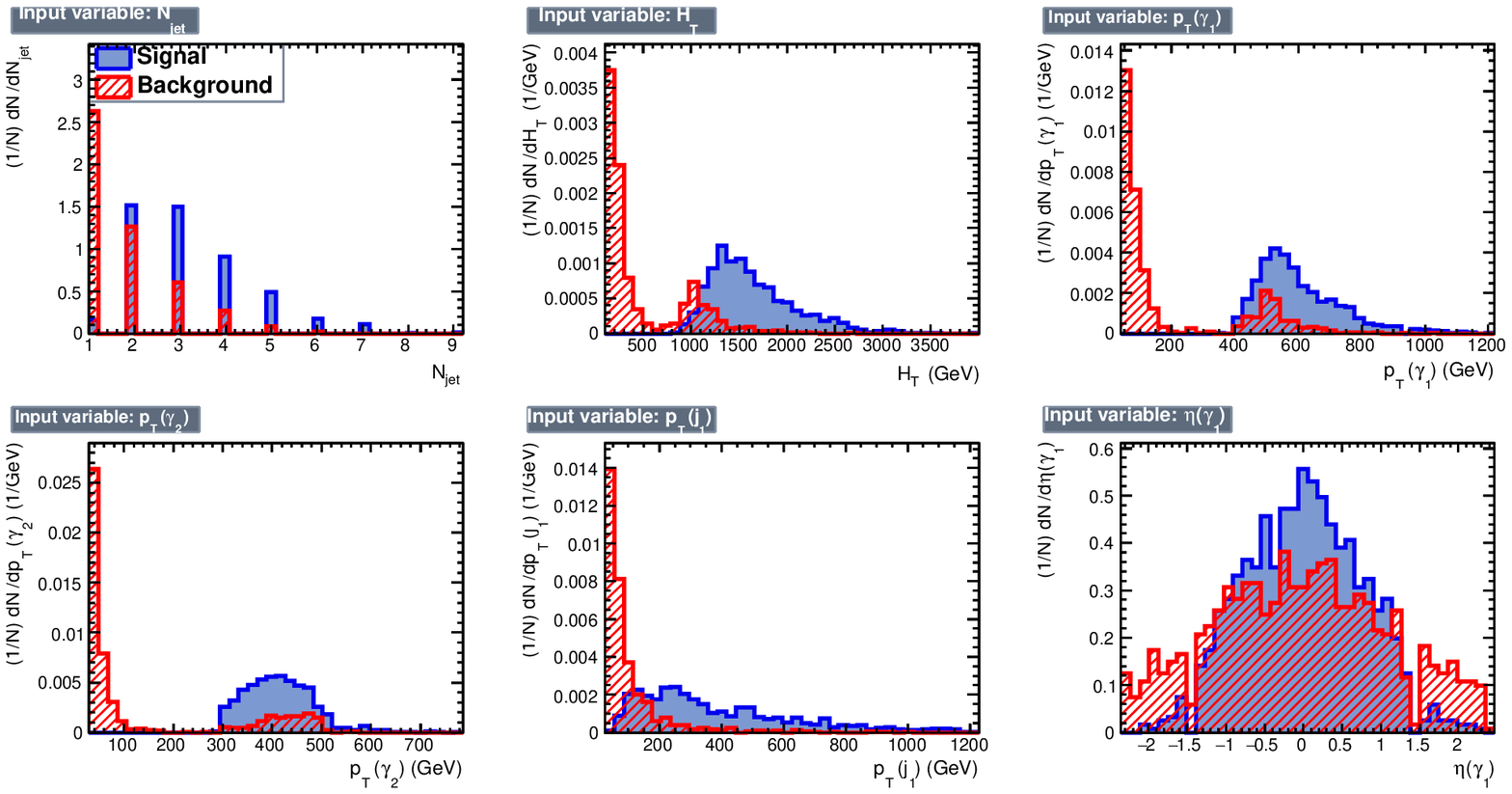}
\caption{Normalized distributions of a few sample input variables for the signal ($WW$ fusion with $M_{\Phi}=1$ TeV) and
background at the LHC ($\sqrt{s}=13$ TeV). These distributions are obtained 
by applying the selection cuts defined in Section~\ref{ssec:exclu}.}
\label{fig:varww}
\end{figure}

\begin{figure}[h!]
\includegraphics[scale=0.35]{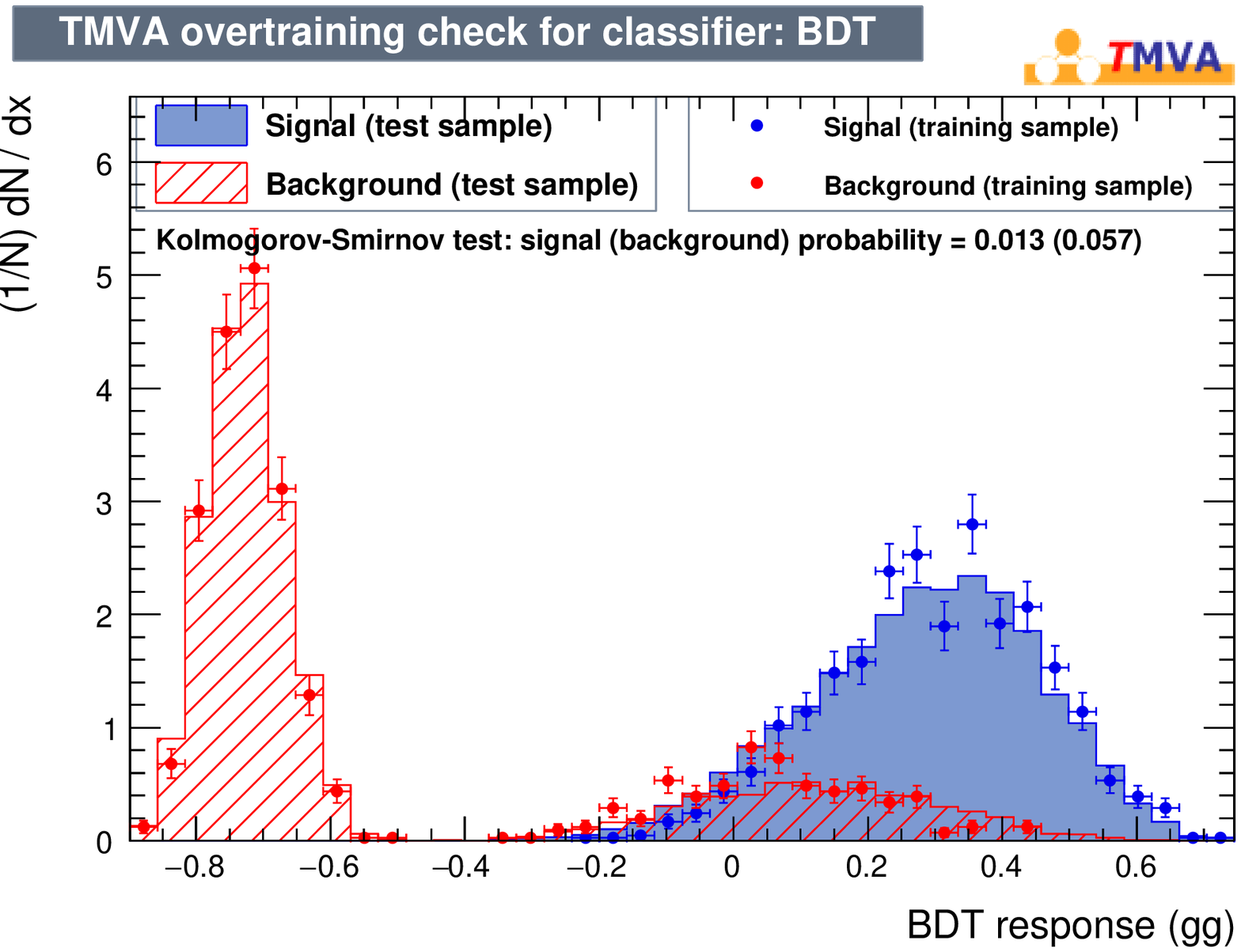}
\includegraphics[scale=0.35]{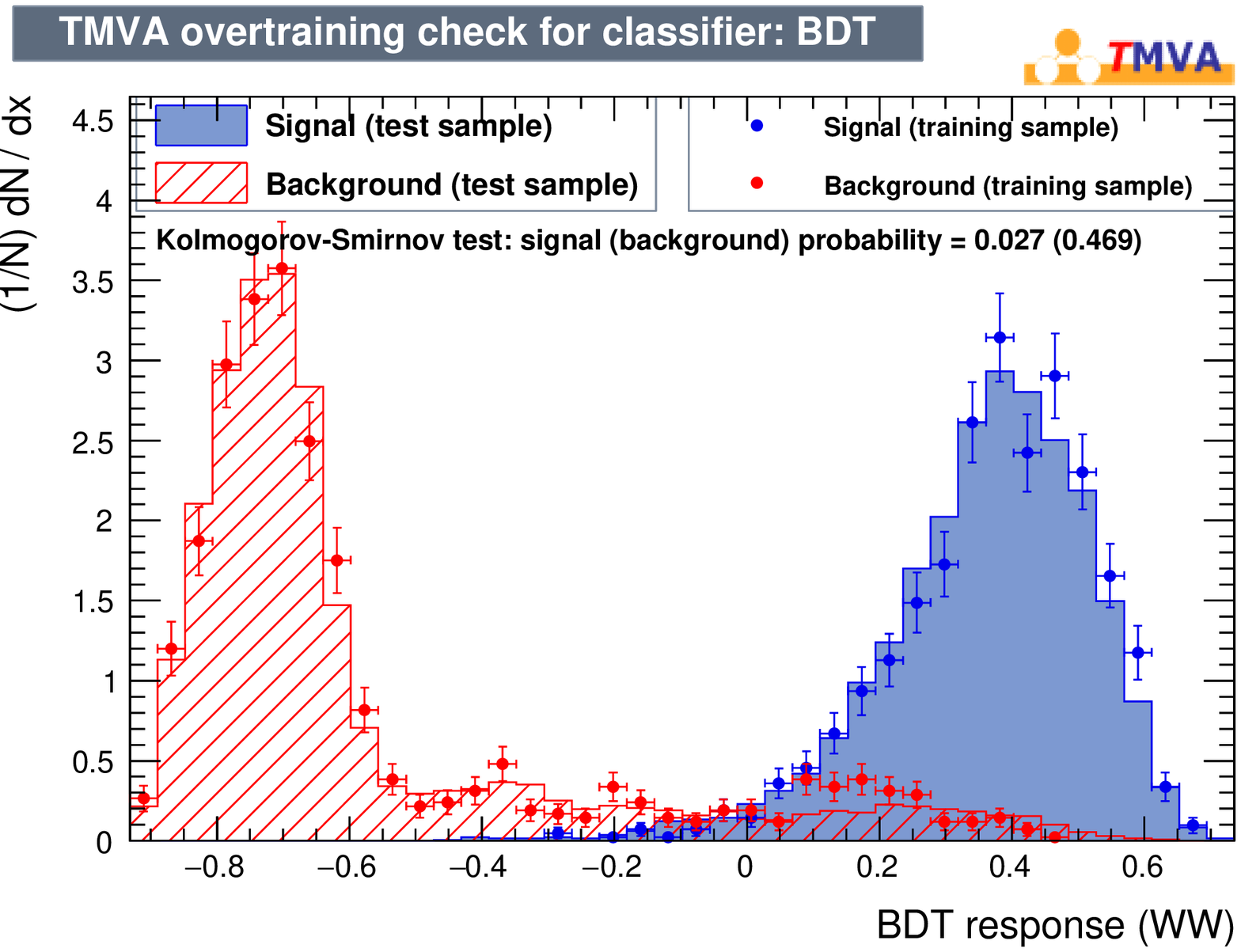}
\caption{The BDT response for the signal (left panel: $gg$ fusion and right panel: $WW$ fusion) and 
the background for $M_{\Phi}=1$ TeV at the $\sqrt{s}=13$ TeV LHC.}
\label{fig:BDTggww}
\end{figure}

For the interested readers, we show here the distributions of some input variables for the signal and the
background. The signal in Figs.~\ref{fig:vargg} and \ref{fig:varww} are for the $gg$ and the $WW$ fusion production modes, respectively. These distributions are obtained by applying the selection cuts defined in Section~\ref{ssec:exclu}. 
The BDT responses for these two production modes with background are presented in Fig.~\ref{fig:BDTggww}.
We observe that the signal and the background distributions are very different in nature and, therefore, one could use a MVA to isolate the signal 
from the background. After filtering out the signal events from the background, one can use
our method to identify the underlying production mechanism.
It is expected that the signal distributions deviate
more and more from the background as we increase the resonance mass. Therefore, isolation of the signal
from the background becomes easier for heavier resonances. One can, therefore, tune MVA
for lower masses and use the same optimized analysis for higher masses, for simplicity.

Notice that there is a second bump around 500 GeV and in the range $300-500$ GeV in the background $p_T(\gm_1)$ 
and $p_T(\gm_2)$ distributions respectively. Similarly, there is a second bump in the background $H_T$ distribution around 1000 GeV
(this is expected since $H_T$ is correlated with the transverse momenta of the photons). This unusual shape of these distributions also leads to the bimodal nature of the 
background BDT responses in Fig.~\ref{fig:BDTggww}. The origin of these peculiar second bumps in the
background distributions is due to the selection cuts $E_T(\gm_1) > 0.4M(\gm_1,\gm_2)$
and $E_T(\gm_2) > 0.3M(\gm_1,\gm_2)$ used to obtained these plots. We have confirmed that these bumps go away with the 
removal of the above-mentioned correlated cuts.


\end{document}